\documentclass[aps,prb,superscriptaddress,twocolumn]{revtex4}

\usepackage{amsmath}
\usepackage[pdftex]{graphicx}
\usepackage{epstopdf}
\usepackage{xcolor}
\usepackage{booktabs}
\usepackage[abs]{overpic}

\def \degree{{^{\circ}}}

\def \bn{\begin{align}}
\def \en{\end{align}}
\def \be{\begin{equation}}
\def \ee{\end{equation}}
\def \bea{\begin{eqnarray}}
\def \eea{\end{eqnarray}}
\def \ba{\begin{array}}
\def \ea{\end{array}}

\def \av#1{{\langle#1\rangle}}

\def \bra#1{{\langle#1|}}
\def \ket#1{{|#1\rangle}}

\def \e{{\epsilon}}
\def \a{{\alpha}}

\def \w{{\omega}}

\def \D{{\Delta}}

\def \yd{^\dagger}

\def \etal {{\it et al. }}

\def \nn{{\nonumber}}

\usepackage{ulem}

\newcommand{\braket}[2]{{\langle  #1 | #2 \rangle}}
\newcommand{\braketop}[3]{\langle #1 | #2 | #3 \rangle}

\newcommand{\etahat}[0]{\hat{\eta}}

\renewcommand{\vr}[0]{{\bf r}}

\renewcommand{\epsilon}{\varepsilon}

\begin{document}

\title{Friedel Oscillations as a Probe of Fermionic Quasiparticles}

\author{Emanuele G. Dalla Torre}
\affiliation{Department of Physics, Bar Ilan University, Ramat Gan 5290002, Israel}
\affiliation{Department of Physics, Harvard University, Cambridge, MA 02138, U.S.A.}
\author{David Benjamin}
\affiliation{Department of Physics, Harvard University, Cambridge, MA 02138, U.S.A.}
\author{Yang He}
\affiliation{Department of Physics, Harvard University, Cambridge, MA 02138, U.S.A.}
\author{David Dentelski}
\affiliation{Department of Physics, Bar Ilan University, Ramat Gan 5290002, Israel}
\author{Eugene Demler}
\affiliation{Department of Physics, Harvard University, Cambridge, MA 02138, U.S.A.}

\begin{abstract} 
When immersed in a see of electrons, local impurities give rise to density modulations known as Friedel oscillations. In spite of the generality of this phenomenon, the exact shape of these modulations is usually computed only for non-interacting electrons with a quadratic dispersion relation. 
In actual materials, Friedel oscillations are a viable way to access the properties of electronic quasiparticles, including their dispersion relation, lifetime, and pairing. In this work we analyze the signatures of Friedel oscillations in STM and X-ray scattering experiments, focusing on the concrete example of cuprates superconductors. We identify signatures of Friedel oscillations seeded by impurities and vortexes, and explain experimental observations that have been previously attributed to a competing charge order. 
\end{abstract}

\maketitle

\tableofcontents

\section{Introduction}
In the study of strongly correlated materials, one of the common themes is the appearance of competing and/or intertwined orders (See for example Refs.~[\onlinecite{kivelson03,sachdev04_competing,davis13_concepts}] for an review). Recently this subject received considerable attention due to the experimental observation of incommensurate charge modulations that coexist with high-temperature superconductivity in cuprates\cite{ghiringhelli12,chang12,comin13,yazdani14,shen14,tabis14,fujita14,hashimoto14,hamidian15_atomic,hamidian15_magnetic}. These modulations were originally observed on the surface of BSCCO via scanning tunneling spectroscopy (STM) \cite{hoffman02B,kapitulnik03, davis04,yazdani04}, and recently found in the bulk of YBCO and other materials by X-ray scattering experiments \cite{ghiringhelli12,chang12,comin13,yazdani14,shen14,tabis14}. It is commonly believed that these modulations are adiabatically connected to the long-ranged charge order  observed at large magnetic fields by quantum oscillations \cite{doiron07,barivsic13} and nuclear magnetic resonance (NMR) \cite{wu11}. In our earlier work \cite{dallatorre15} we proposed an alternative interpretation: the modulations observed at small magnetic fields can be understood as Friedel oscillations due to the scattering of quasiparticles with a short lifetime, rather than as the evidence of a competing order.

Friedel oscillations are density modulations generated by local impurities acting on mobile charges, such as electrons in a metal. At the lowest order of perturbation theory, these modulations are proportional to the static density-density response function of the unperturbed (homogeneous) system. For free electrons in three dimensions, this function can be analytically computed and its Fourier transform is peaked at twice the Fermi wavevector (see Ref.~[\onlinecite{mihaila11}] for a review). As a consequence, Friedel oscillations can be exploited to directly measure the electron density \cite{sprunger97_giant,petersen98_direct}. In more complex  materials the shape of Friedel oscillations is determined by the band structure of fermionic quasiparticles, their lifetime, and the presence of a pairing gap. We suggest that the observation of Friedel oscillations is therefore a viable tool for studying the properties of quasiparticles in strongly-correlated materials.

In this paper we theoretically analyze signatures of Friedel oscillations in X-ray and STM experiments, focusing on the specific example of superconducting cuprates. The band structure of these materials has been extensively studied by angle-resolved photoemission spectroscopy (ARPES). In the normal phase, these materials have a single Fermi surface, whose phenomenological parameters are well known. This information allows us to make quantitative predictions for the expected shape of the Freidel oscillations. At low temperatures, the presence of a superconducting gap challenges the main assumption of Friedel oscillations, namely the presence of an underlying Fermi liquid. As we will see below, Friedel oscillations can occur in a paired state as well, with some important differences. In particular, in a superconductor, Friedel oscillations can be seeded by both local modulations of the chemical potential (as in normal metal) and local modulations of the pairing gap.

In the following we fist present the theoretical framework used to analyze Friedel oscillations and then discuss its implications for recent experiments in cuprates. The appendices explain in detail several technical details of the calculations and in particular: how to perform an ensemble average over static impurities (Sec.~\ref{app:impurities}); how to derive the formalism of resonant elastic scattering using Green's functions (Sec.~\ref{app:REXS}) and how to relate it to the Lindhard susceptibility (Sec.~\ref{app:REXS2}); how to quantify the effects of the spin-orbit coupling (Sec.~\ref{app:spinorbit}) and of the light-field polarization (Sec.~\ref{app:polarization}).

\section{Theoretical framework}
\label{sec:theory123}

\subsection{Lindhard susceptibility}
\label{sec:theory0}
We open our theoretical description of Friedel oscillations by considering (hard) X-ray scattering experiments. This probe was used to detect Friedel oscillations in vanadium-doped blue bronze\cite{rouziere00}, a charge density wave (CDW) material. For this material, accurate X-ray scattering experiments revealed two distinct incommensurate diffraction peaks. These peaks where respectively identified with the CDW wave-vector, and with Friedel oscillations at twice the Fermi wavevector.  In a typical X-ray experiment the intensity of the scattered light is proportional to the zero-frequency density-density response function \cite{ashkroft_book}
\be \chi({\bf q})=\int_0^\infty dt~\int dx~e^{i{\bf q}\cdot{\bf x}} \langle\Big[\rho({\bf x},t),\rho({\bf x},0)\Big]\rangle, \label{eq:chi0}\ee
where $[\cdot,\cdot]$ is the commutation relation and $\rho(x,t)=\psi\yd(x,t)\psi(x,t)$ is the charge density. For quasiparticles with a dispersion relation $\epsilon_k$ and a finite lifetime $\Gamma$, Eq.~\eqref{eq:chi0} becomes
\be
\chi(q)=\sum_{\bf k}~ \frac{n_{\bf k}-n_{\bf k+q}}{\epsilon_{\bf k} - \epsilon_{\bf k+q} + 2 i \Gamma} \label{eq:lindhard}\;.\ee
where $n_{\bf k}=[1+\exp((\epsilon_k-\mu)/T)]^{-1}$ is the Fermi-Dirac distribution function, and $T$ the temperature. By neglecting interactions between quasiparticles, Eq.~(\ref{eq:lindhard}) disregards possible collective modes such as spin waves and paramagnons.
In the case of free electrons (with $\epsilon_{\bf k}=k^2/2m$ and $\Gamma\to0^+$) and at $T=0$, Eq.~(\ref{eq:lindhard}) can be evaluated analytically\cite{mihaila11}. In two dimensions $\chi({\bf q})$ is momentum-independent for $q<2k_F$, and decays algebraically for $q>2k_F$, where $k_F$ is the Fermi momentum \footnote{Note that current experiments would detect this situation as a pronounced peak at $q=2k_F$ due to the common practice of ``background subtraction''.}. In actual materials, the dispersion relation is more complex and an exact analytical evaluation of Eq.~(\ref{eq:lindhard}) is generically not possible. We therefore resort to a numerical evaluation of this expression. As we will see in Sec.~\ref{sec:lindhard}, this calculation leads to sharp peaks in $\chi(q)$. When such peaks are observed in experiments, they may be interpreted as evidence of static charge density waves (CDW).


\subsection{Local density of states}
\label{sec:theory}
In contrast to hard X-ray measurements, STM and REXS temporarily change the number of electrons in the conduction band and couple to the density of states, rather than to the density-density response function\cite{abbamonte12}. Specifically, at zero temperature the STM differential conductivity $dI({\bf r})/dV$ is proportional to the local density of states $g({\bf r},\w=V)$, given by the imaginary part of the retarded Green's function:
\begin{align}
g({\bf r},\w) &= {\rm Im}[G({\bf r},\omega)]\\
& ={\rm  Im}\left[ \int_0^\infty dt~e^{-i\omega (t-t')}\av{[\psi\yd({\bf r},t),\psi(\bf{ r},t')]}\right]\label{eq:grw}\;.
\end{align}
For disordered materials, $g({\bf r},\w)$ varies in space and is in general unpredictable. It is therefore common to compute the two-dimensional Fourier transform of the signal at a fixed voltage\cite{petersen98_direct}
\be g({\bf q},\omega) = \int d^dr~e^{i {\bf q}\cdot{\bf r}} ~ g({\bf r},\omega)\;.\label{eq:grq} \ee
As we will explain in detail below, the absolute value of $g({\bf q},\omega)$ depends on the types of scatterers present in the material, but not on their position (assuming that the sample is large enough to enable self-averaging of the scatterers' position).

\begin{figure}[t]
\includegraphics[scale=0.3]{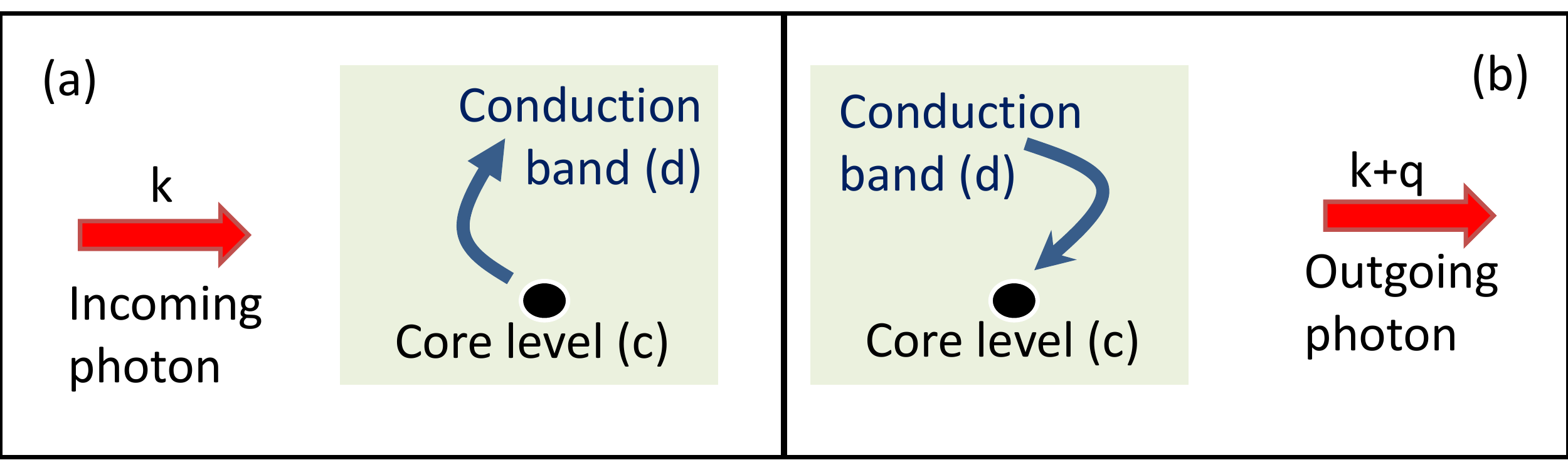}
\caption{Schematic diagram of a typical REXS experiment, in which X-ray photons scatter electrons from a core level to the conductions band (a), and viceversa (b).}
\label{fig:REXS}
\end{figure}

Resonant elastic X-ray scattering (REXS) offers an alternative way to  measure the local density of states, $g({\bf q},\omega)$. As pointed out by Abbamonte \etal \cite{abbamonte12}, STM and REXS describe analogous processes: in STM electrons tunnel to the sample's conduction band from an atomic-size tip, while in REXS they are coherently pumped from a local core level (see Fig.~\ref{fig:REXS}). Based on this analogy, Abbamonted \etal modeled the intensity of the REXS signal (at zero temperature) by
\be I_{\rm REXS}({\bf q},\omega) = \left| A \int_{0}^\infty d\w'  G^R_c(\w-\w') g({\bf q},\w')\right|^2\label{eq:REXS} \ee
Here $G^R_c(\w)=[(\w+i\Gamma_c)]^{-1}$ is the retarded Green's function of the core level and $\Gamma_c$ is its lifetime. In Appendix \ref{app:REXS} we provide a derivation of Eq.(\ref{eq:REXS}) based on the Keldysh Green's function formalism, which allows to extend this expression to finite temperatures. Furthermore, in Appendix \ref{app:REXS2} we show that in the limit of non resonant scattering ($\Gamma_c\to \infty$) from a Fermi sea, Eq.~(\ref{eq:REXS}) reduces to the Lindhard susceptibility (\ref{eq:lindhard}). Notably, Eq.~(\ref{eq:REXS}) neglects the effects of the core-hole potential on the evolution of the conduction band. This effect is fundamental to understand resonant inelastic scattering (RIXS)\cite{ament11} processes, but can probably be neglected in the case of REXS.

The prefactor $A$ in Eq.~(\ref{eq:REXS}) describes the transition amplitude for the excitation of a single core hole. As shown in Appendix \ref{app:spinorbit}, this quantity does not depend on the details of the core orbital and, in particular, is unaffected by the spin-orbit coupling. In the absence of magnetic impurities, the dipole approximation results into 
\be A \propto \braketop{d}{ \left( \etahat_i \cdot \vr \right) \left( \etahat^*_o \cdot \vr \right)
 }{d};,\label{eq:polarization2}\ee
where $\ket{d}$ denotes the orbital wavefunction of the electrons forming the conduction band, $\vr$ is displacement vector in this state, and $\etahat_{i/o}$ is the polarization of the incoming/outgoing photon. Eq.~(\ref{eq:polarization2}) is used in Appendix \ref{app:polarization} to compare the theoretical predictions of the present single-band model with the experimental results of Ref.~[\onlinecite{comin14}]. 

\subsection{Wannier functions and Bragg peaks}
\label{sec:theory1bis}
We now consider the effects of non trivial Wannier functions on the Fourier-transformed local density of states $g({\bf q},\omega)$. To achieve this goal, we first express 
Eq.~(\ref{eq:grq}) in terms of the Fourier-transformed fermionic operators $\psi_k(t)=\int d^d r~ e^{i{\bf k}\cdot{\bf r}}\psi({\bf r},t)$ and their retarded Greens function $G({\bf k},\bf{k+q},t)=\int_0^\infty~dt~e^{-i\omega t} \av{[\psi({\bf k},t),\psi\yd({\bf k+q},t)]}$ as
\be g({\bf q},\omega)= {\rm Im}\sum_{\bf k}~G({\bf k},{\bf k+q},\omega)\;,\label{eq:gqw} \ee
To derive Eq.~\eqref{eq:gqw} we assumed the system to be symmetric under ${\bf r}\to-{\bf r}$. This symmetry allowed us to invert the order of the $\int d^dk$ and ${\rm Im}$ operators (see SI-2 of Ref.~[\onlinecite{dallatorre13}]). This assumption is valid for example in the presence of a single scatterer at the origin of the axis. In the presence of several scatterers at random locations the present analysis applies to the absolute value of the measured quantity (see Appendix ~\ref{app:impurities}). 

For a single-band model, the operator $\psi({\bf r},t)$ is related to the annihilation of an electron (quasiparticle) on a single site, $c_i$, through the Wannier function $W({\bf r-r}_i)$,
\be \psi({\bf r},t)=\sum_i W({\bf r}-{\bf r_i})c_i\label{eq:Wannier}\;. \ee
Combining Eqs. (\ref{eq:gqw}) and (\ref{eq:Wannier}) we arrive to the expression \cite{podolsky03,dallatorre13,choubey14,kreisel15}
\be g({\bf q},\w)= {\rm Im}\sum_{\bf k}~ W^*_{\bf k} G_{\rm lattice}({\bf k},{\bf k+q},\omega)~W_{{\bf k+q}}\;, \ee
where $W({\bf k})=\int d^dr e^{i{\bf k}\cdot{\bf r}}W({\bf r})$, $G_{\rm lattice}({\bf k},{\bf k+q},t)=\int_0^\infty~dt~e^{-i\omega t} \av{[c({\bf k},t),c\yd({\bf k+q}t)]}$, and $c_{\bf k}=\sum_i e^{i {\bf k}\cdot{ x}_{i}} c_i$. Note that by definition, $G_{\rm lattice}({\bf k},{\bf k+q},\omega)$ is  a periodic function of ${\bf k}$ and ${\bf q}$ with a period given by the Bravais lattice vectors, ${\bf G}$. In the absence of impurities $G_{\rm lattice}({\bf k},{\bf k+q},\omega) = G({\bf k},\omega)\sum_{{\bf G}}\delta({\bf q}-{\bf G})$, where $G({\bf k},\omega)=G({\bf k},{\bf k},\omega)$. This expression gives rise to well-defined Bragg peaks in $g({\bf q},\omega)$, whose intensity is determined by the width of the Wannier function.

Cuprates posses a non-trivial Wannier function with $d$-wave symmetry \cite{zhang88}. As explained in Ref.~[\onlinecite{dallatorre16}], this shape affects the phase and intensity of the Fourier-transformed STM signal at large wavevector (beyond the central Brillouin zone), leading to subtle correlations that were interpreted as evidence of a competing order with $d$-form factor\cite{fujita14,hashimoto14,hamidian15_atomic,hamidian15_magnetic}. In this paper we focus on the central Brillouin zone, where the precise shape of the Wannier function is not very important. For convenience, we then approximate $W({\bf k})$ as a Gaussian wavefunction with width $\sigma_k = 1.8(2\pi/a)$, 
where $a$ is the lattice constant.

\begin{figure}
\includegraphics[scale=0.4]{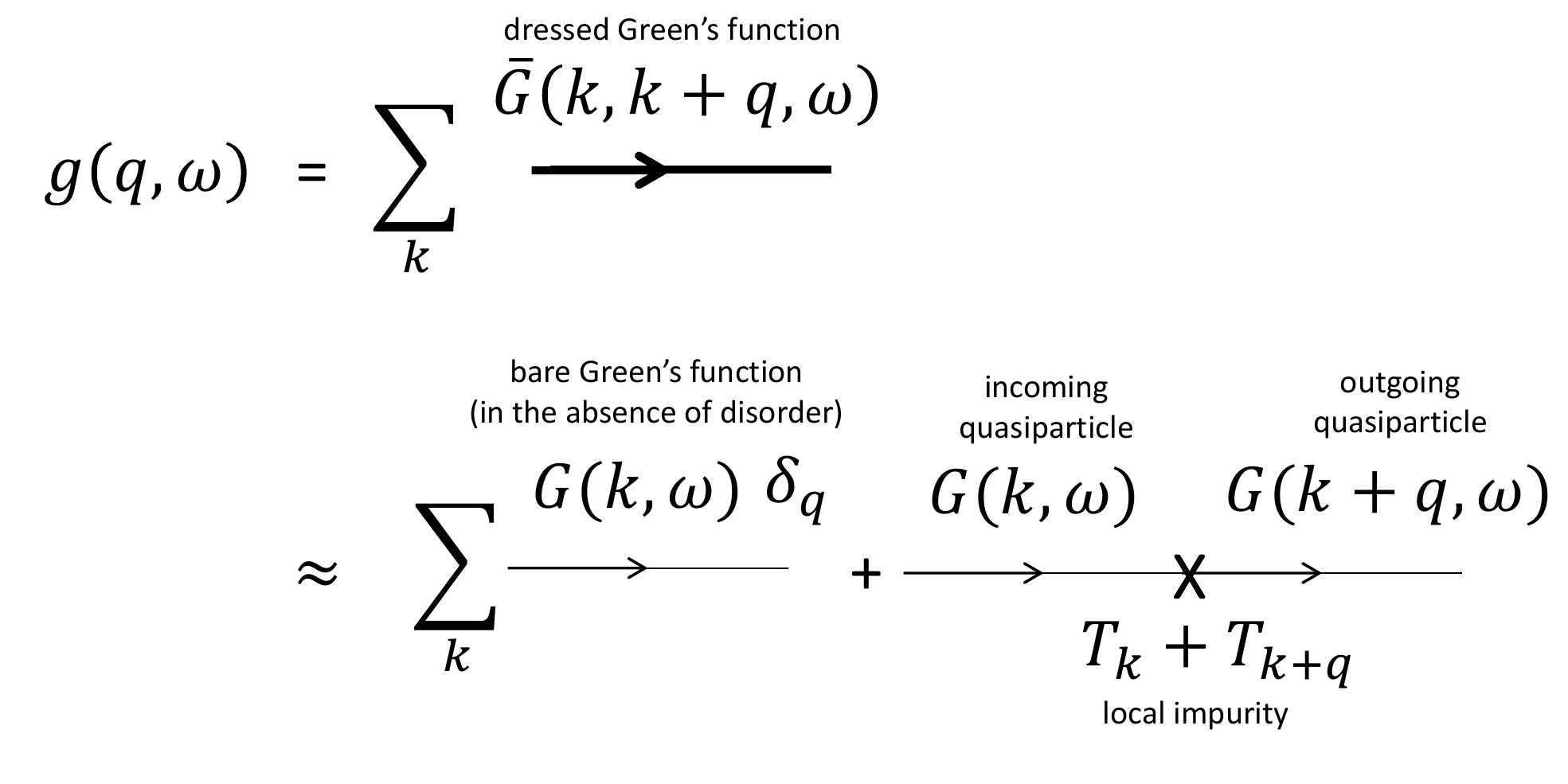}
\caption{Schematic representation of the diagrams considered in the present analysis. Our approach is based on the first-order perturbation theory in the strength of the disorder and does not include the effects of interactions among quasiparticles. The function $g(q,\omega)$ is the Fourier-transformed local density of states.}
\label{fig:born}
\end{figure}

\subsection{Scattering from local impurities}
\label{sec:theory2}

In actual materials, as a consequence of disorder, the Fourier-transformed density of states $g({\bf q},\omega$) is non zero even for wave-vectors that do not correspond to a lattice vector. 
Performing a first-order perturbation theory in the scattering potential (Born approximation) one finds\cite{nowadnick12}
\begin{align} G({\bf k},{\bf k+q},\omega) &= G({\bf k},\omega)\sum_{{\bf G}}\delta({\bf q}-{\bf G})\nonumber \\
&+ G({\bf k},\omega) T({\bf k},{\bf q}) G({\bf k+q},\omega)\;, \end{align}
where $G({\bf k},\omega)=G({\bf k},{\bf k},\omega)$, and $T({\bf k},{\bf q})$ describes the scattering of quasiparticles from momentum ${\bf k}$ to momentum ${\bf k+q}$.

One of the main goals of this paper is to consider the effects of different types of impurities, defined through their scattering matrices $T$. We consider here only perturbations that are static and quadratic in the quasiparticles' creation and anihilation operators. Any such perturbation can be described by the Hamiltonian 
\be H_{\rm pert} = V_{i,j} c\yd_i c_j =\sum_{\bf k,q} T({\bf k,q})c\yd_{\bf k} c_{\bf k+q}\label{eq:Tkq}\; \ee
where $T({\bf k,\bf q})= \sum_{i,j}V_{i,j} (e^{i {\bf q}\cdot {\bf x}_j}e^{i {\bf k}\cdot ({\bf x}_i-{\bf x}_j)} + e^{i {\bf q}\cdot {\bf x}_i}e^{i {\bf k}\cdot ({\bf x}_j-{\bf x}_i)})$. 

If the scatterer acts on a single site (or on the bonds linked to a single site), the scattering amplitude is given by the sum of two terms, which depend respectively on the momentum of the incoming and outgoing quasiparticles only:  $T({\bf k},{\bf q})=T_{\bf k}+T_{\bf k+q}$ . Combining this expression with Eq.~(\ref{eq:gqw}) we find
\begin{align} & g({\bf q},\w) =\sum_{\bf k} {\rm Im} \left[W^*_{\bf k}G({\bf k},\omega)W_{\bf k+q}\right]\sum_{{\bf G}}\delta({\bf q}-{\bf G})\nonumber \\
& + \sum_{\bf k}~{\rm Im}\Big[ W^*_{\bf k} G({\bf k},\w)(T_{\bf k}+T_{\bf k+q}) G({\bf k+q},\w) W_{\bf k+q}\Big]\label{eq:rhoqw2}\;.\end{align}
Eq.~(\ref{eq:rhoqw2}) is at the basis of the present analysis: the first line corresponds to the density of states in an ideal lattice, while the second line describes the effects of the impurities (see also Fig.~\ref{fig:born}). In what follows we will mainly consider this latter contribution.

\subsection{Impurities in a paired state}
\label{sec:theory3}

The above-mentioned formalism can be easily extended to include the effects of a spectral gap. For concreteness, we describe underdoped cuprates in terms of a single spectral gap, the paring gap $\Delta$. Following Ref.~[\onlinecite{dallatorre13}], we propose that the second energy scale observed in many experiments corresponds to the quasiparticle's lifetime $\Gamma$, rather than to a distinct (competing) gap. This would explain the uncertainty in determining the precise value of the gap in underdoped samples, varying roughly between $\Delta-\Gamma$ and $\Delta+\Gamma$, depending on the type of experiment \cite{hufner08}. 
The relatively-large value of $\Gamma$\cite{alldredge08} in underdoped cuparates might be related to enhanced phase fluctuations, which lead to a loss of global phase coherence at the critical temperature $T_c$ (see for example Refs. [\onlinecite{uemura89,doniach90,randeria92,emery95,chen1998,franz1998,deutscher1999,kwon1999,carlson1999}]). Notably, the density of states is a gauge-invariant object and, as such, depends only on the amplitude of $\Delta$ but not on its phase. 
For simplicity, we assume the pairing gap to have a pure d-wave form, $\Delta_{\bf k} = \Delta_0/2(\cos k_x - \cos k_y)$, although this assumption has little effect on the final result. Conforming to the Born approximation, we assume $\Delta_0$ to be homogeneous over the sample and independent on the impurities. 

In the presence of a pairing gap, quasiparticles are conveniently represented as $2\times2$ matrices in Nambu space (whose two entries are respectively particles and holes). In this notation the retarded Green's function of a  quasiparticle with momentum ${\bf k}$ and energy $\omega$ is given by
\be G^{-1}({\bf k},\w) = \left(\ba{c c}\w - \e_{\bf k} +\mu + i \Gamma & \D_{\bf k} \\ \D_{\bf -k} & \w + \e_{\bf -k} -\mu + i \Gamma \ea\right)\;.~\label{eq:G0}\ee
The dispersion relation $\epsilon_{\bf k}$, the pairing gap $\Delta_0$, and the quasiparticles lifetime $\Gamma$ relevant to superconducting cuprates are provided in Sec.~\ref{sec:experiment} and in Table \ref{table1}.

Static and quadratic perturbations can be divided into two main categories, CDW and PW, depending on whether they conserve the total number of quasiparticles ($\sim c\yd_i c_j$, see Eq.~(\ref{eq:Tkq}))), or not ($\sim c_i c_j + c\yd_i c\yd_j$), often referred to respectively as diagonal and off-diagonal. In this paper we restrict our analysis to three specific types of impurities: two of them have a simple physical interpretation and correspond to local modulations of the chemical potential (sCDW) and of the pairing gap (dPW). The third type (dCDW) corresponds to a local modulation of the intra-unit-cell nematic order \cite{lawler10} and has d-wave symmetry. These three types of impurities correspond to the real-space Hamiltonians $H^{sCDW} =  c_{0,0}\yd c_{0,0}$, $H^{dCDW} = (c_{0,0}\yd c_{1,0}+c_{0,0}\yd c_{-1,0}) - (c_{0,0}\yd c_{0,1}+c_{0,0}\yd c_{0,-1})+ {\rm H.c.} $, $H^{dPW} = (c_{0,0}\yd c\yd_{1,0}+c\yd_{0,0} c\yd_{-1,0}) - (c\yd_{0,0} c\yd_{0,1}+c\yd_{0,0} c\yd_{0,-1}) +{\rm H.c.}$. The associated scattering matrices to be used in Eq.~(\ref{eq:rhoqw2}) are
\begin{align}
T^{sCDW}_{\bf k} &= \left(\ba {c c}1 & 0 \\ 0 & -1\ea\right),\nn\\
T^{dCDW}_{\bf k} &= \left(\ba {c c} d_{\bf k} &0 \\ 0& -d_{\bf k}\ea\right)\nn,\\
T^{dPW}_{\bf k} &= \left(\ba {c c} 0 & d_{\bf k} \\ d_{\bf k} & 0\ea\right)\label{eq:T3}
\end{align}
where $d_{\bf k}=\cos k_x -\cos k_y$. 
The Fourier transformed density of states $g({\bf q},\omega)$ is obtained by numerically integrating Eq.~(\ref{eq:rhoqw2}), with $G({\bf k},\omega)$ and $T_{\bf k}$ respectively defined by Eqs. (\ref{eq:G0}) and (\ref{eq:T3}). As we will see, a comparison between the resulting plots and the experimental findings suggests a coexistence of sCDW and dPW, but rules out the presence of dCDW local modulations.


\section{Experiments on Cuprates}

\label{sec:experiment}

\subsection{X-ray: effects of the band structure}
\label{sec:lindhard}

As mentioned in Sec.~\ref{sec:theory0}, X-ray experiments couple to the density-density response function and can be used to directly measure Friedel oscillations. To employ the Lindhard formula (\ref{eq:lindhard}) it is necessary to know the dispersion relation $\epsilon_{\bf k}$, the chemical potential $\mu$, and the quasiparticles' lifetime $\Gamma$. In the case of superconducting cuprates, these parameters can be directly read from accurate angle-resolved photoemission spectroscopy (ARPES) experiments. Following the common approach, we assume electrons to move within isolated CuO planes and map the conduction band in terms of the two-dimensional dispersion relation
\begin{align}
& \epsilon_k = \frac{t_0}2\left(\cos k_x +\cos k_y\right) + t_1 \cos k_x \cos k_y \nonumber\\
&+ \frac{t_2}2 \left(\cos k_x +\cos k_y\right) + \frac{t_3}2 \left(\cos 2 k_x \cos k_y +\cos k_x \cos 2 k_y\right)\nonumber \\
& + t_4 \cos 2 k_x \cos 2 k_y + \frac{t_5} 2\left(\cos 2k_x\cos k_x+\cos2k_y\cos k_y\right)\label{eq:Ek}\;.
\end{align}
In this work we specifically refer to three distinct compounds:  Bi$_2$Sr$_{2−x}$La$_x$CuO$_{6+\delta}$ (Bi2201), Bi$_{2}$Sr$_{2}$CaCu$_{2}$O$_{8+\delta}$ (Bi2212), 
$\mathrm{YBa_2Cu_3O_{7-x}}$ (Y123),
whose band structure were experimentally determined by  Norman \etal \cite{campuzano95}, Schabel \etal \cite{shen98}, Pasani \& Atkinson \cite{pasani10}, King \etal \cite{king11}. The relevant parameters $t_0-t_6$ are reproduced in Table \ref{table1}. Note that Y123 material has inequivalent bonding (B) and antibonding (A) bands: Table \ref{table1} refers only to the former one.

\begin{figure}
\centering
\begin{overpic}[scale=0.9,unit=1mm]{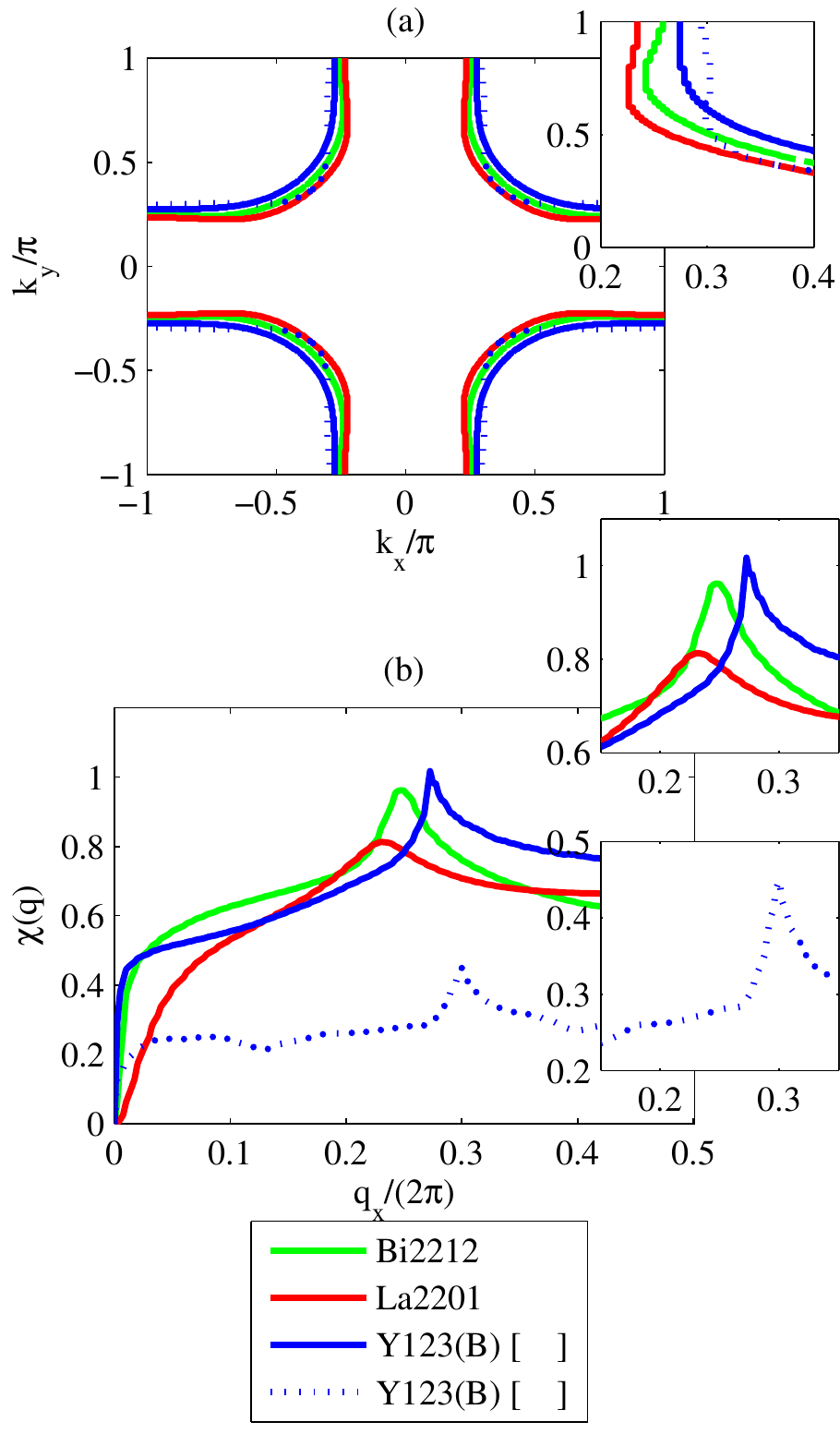}
\put(51.7,7){\Large \cite{pasani10}}
\put(51.7,2){\Large \cite{shen98}}
\end{overpic}

\caption{(a) Fermi surfaces resulting from the band structures listed in Table \ref{table1}. (b) Lindhard response, Eq.(\ref{eq:lindhard}), along the line ${\bf q}=(2\pi/a)\times(q,0)$ for the same materials.}
\label{fig:lindhard}
\end{figure}

The chemical potential $\mu$ is uniquely determined by the charge doping through the Luttinger count. Following the common convention, we denote by $p$ the density of additional holes with respect to half filling:
\be p= 2x - 1 ~,~~{\rm where}~~ x =\frac{\sum_k n_k}{ \sum_k}\;,\ee and $k$ runs over the Brillouin zone. In the case of Y123, we identify the nominal doping with the algebraic average of the doping in the bonding and antiboding bands, $p=(p_A+p_B)/2$. The resulting Fermi surfaces are plotted in Fig.~\ref{fig:lindhard}(a). As shown in the inset, the Fermi surfaces are not circular, and display a significant amount of nesting at the antinodes \footnote{The antinodes are here defined as the points at which the Fermi surface reaches the boundaries of the first Brillouin Zone}.

\begin{table}

\begin{tabular}{| c |r| r |r r|}
\hline
& Bi2212 & Bi2201 & \multicolumn{2}{c|}{Y123(B)}\\Band Structure& [\onlinecite{campuzano95}] & [\onlinecite{king11}] & [\onlinecite{pasani10}] & [\onlinecite{shen98}]
\\ \hline
$  \mu $   & 0.0234 & -0.148 & -0.03 & -0.1256 
\\
$ t_0 $ &  -0.5951& -0.5280  & - 0.42 &-1.1259
\\ 
$ t_1 $ & 0.1636 & 0.2438 &  0.1163 & 0.5540
\\ 
$t_2 $  & -0.0519 & -0.0429  & - 0.0983 &-0.1774
\\ 
$ t_3 $ & -0.1117 & -0.0281 & -0.353 &-0.0701
\\  
$ t_4 $ & 0.0510  & -0.0140 & 0 & 0.1286
\\  
$  t_5 $ & 0 & 0 & 0 &-0.1 
\\ \hline
Doping, p &0.04 & 0.11 & \multicolumn{2}{c|}{p$_B$=-0.04} \\
 {} & {} & {} &  \multicolumn{2}{c|}{(p$_A$+p$_B$)/2=0.12}
\\ \hline
Lifetime, $\Gamma$ & 0.004 & 0.020 & 0.002 & 0.001 
\\ \hline 
Gap, $\Delta_0$ & 0.040 & 0.080 & 0.030 & 0.030 
\\ \hline
\end{tabular}
\caption{Phenomenological band structures used in this paper. The parameters $t_0-t_5$ correspond to hopping terms in a tight-binding model and are defined in Eq.~\ref{eq:Ek}. With respect to the originally published band structures, the chemical potential has been shifted to achieve the required doping $p$ (through the Luttinger count). Additionally, 
the parameter $t_5$ has been added to the band structure of Y123(B) in order to cure a spurious back-banding of the band structure (See SI-7 of Ref.~[\onlinecite{dallatorre13}]). $\Gamma$ is the quasiparticle lifetime, used in Eqs.~(\ref{eq:lindhard}) and (\ref{eq:G0}), and $\Delta_0$ the zero-temperature pairing gap used in (\ref{eq:G0}). Their value can be read from the voltage dependence of the Fourier-transformed STM signal\cite{alldredge08,dallatorre13,comin13,fujita14,comin14}, or from  ARPES experiments \cite{kondo06,vishik12}.}
\label{table1}
\end{table}

\begin{figure}
\centering
\includegraphics[scale=0.5]{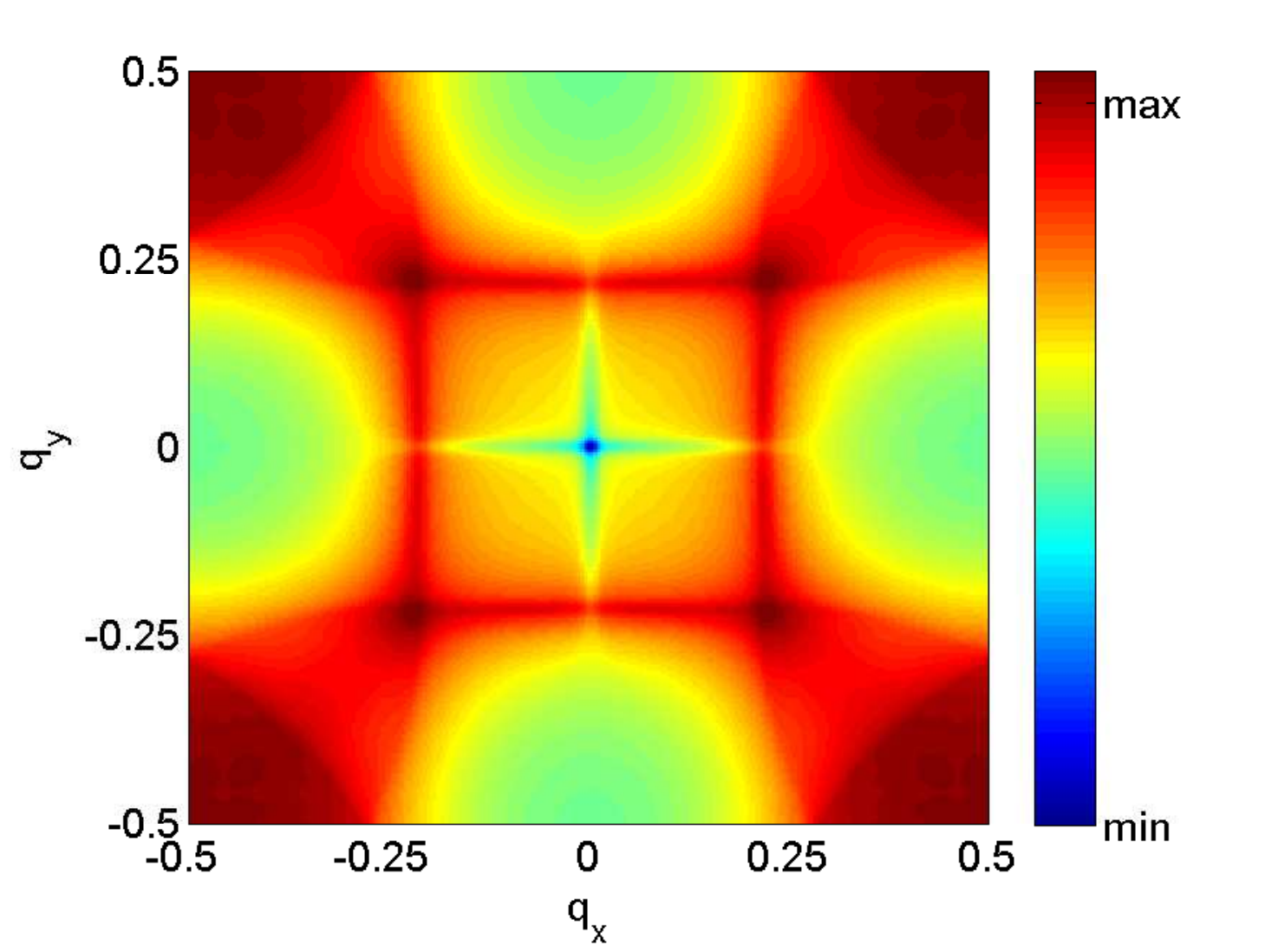}
\caption{Lindhard response, Eq.(\ref{eq:lindhard}), as a function of ${\bf  q} = (2\pi/a)\times(q_x,q_y)$ for Bi2212. The response of the other materials is qualitatively similar (although the peak position is shifted away from $q=0.25$).}\label{fig:lindhardB}
\end{figure}

Using the phenomenological parameters listed in Table~\ref{table1}, we can directly evaluate the Lindhard susceptibility (\ref{eq:lindhard}). Fig.~\ref{fig:lindhard}(b) presents $\chi({\bf q})$ along the direction $(q,0)$, for the three different materials. In all three cases, we observe a pronounced peak at a wave-vector ranging between $0.2$ and $0.3$. The exact position of the peak depends on the choice of the chemical potential, and is roughly equal to the distance between two adjacent antinodes. The width of the peak is of order $0.03-0.1$, leading to a correlation length of about $10-30$ unit cells, or $40-120$A. Its value is mainly determined by the amount of nesting at the antinodes \footnote{Here we use the term ``nesting'' to indicate segments of the Fermi surface that are parallel to each-other, leading to an enhanced scattering at the corresponding wave-length difference}: Among the three materials considered here, the sharpest peak is predicted in Y123, where the amount of nesting  is maximal. In contrast, the Fermi surfaces of Bi2212 and Bi2201 involve a lower level of nesting, resulting in broader peaks. This could explain why, so far, (non-resonant) hard X-ray experiments have revealed Friedel oscillations in Y123 only\cite{chang12}.

The specific choice of the band structure determines the details of the predicted signal. In the case of Y123, Fig.~\ref{fig:lindhard}(b) compares the signal resulting from the band structure of Pasani and Atkinson \cite{pasani10} (continuous blue curve) and of 
Shabel \etal \cite{shen98} (dashed blue curve). As shown in the inset of Fig.~\ref{fig:lindhard}(a), the latter band structure predicts a larger amount of nesting, in agreement with the experiment by Okawa \etal \cite{okawa09}. When used to predict the intensity of the REXS signal, the band structure by Ref.~[\onlinecite{pasani10}] predicts a peak at wavevector $q=0.28$ with width $\delta q\approx 0.05$, while the band structure by Ref.~[\onlinecite{shen98}] predicts a peak at $q=0.3$ with $\delta q\approx 0.03$. For comparison, the experiment of Chang \etal\cite{chang12} shows a peak with maximal intensity at $q=0.31$ and width $\delta q \approx 0.03$, and is found to be in quantitative agreement with the present calculations.

For completeness we mention that the peak observed in X-ray scattering experiments could be additionally enhanced by effected that are not included in the present analysis. In particular, the Linhard formula (\ref{eq:lindhard}) disregards the effects of the electron-phonon coupling. This coupling was instead found to be relatively strong in Y123 at this wavevector, leading to a significant phonon softening\cite{letacon14}. Electron-phonon coupling will generically lead to a sharpening of the X-ray response function, as well as to a renormalization of the position of maximal intensity. It seems plausible that the combination of the band structure of Ref.~[\onlinecite{pasani10}] with electron-phonon coupling could deliver a quantitative agreement with the experiments as good as the one obtained from the band structure of Ref.~[\onlinecite{shen98}]. As a side remark, we also note that the two predicted Lindhard response for Y123   differ by an overall multiplicative factor (see continuous and dashed blue curves in Fig.~\ref{fig:lindhard}(b)). This difference can be traced back to the different bandwidth predicted by the two models ($\sim 0.3eV$ in Ref.~[\onlinecite{pasani10}] and $\sim 1eV$ in Ref.~[\onlinecite{shen98})]. Current experiments involve an unknown normalization factor and are therefore not sufficient to measure the actual value of $\chi(q)$ and distinguish between these two scenarios. Different experiments, and in particular resonant inelastic scattering (RIXS), might be able to fill in this information (see for example Ref.~[\onlinecite{benjamin13}]).




We now consider the full Lindhard susceptibility as a function of the two dimensional wavevector ${\bf q}=(2\pi/a)\times(q_x,q_y)$. Fig.~\ref{fig:lindhardB} represents the results for Bi2212 (whose band structure is known to the highest degree of precision) 
and displays three inequivalent local maxima. The global maximum occurs around the wavector ${\bf q}_{\pi,\pi}=(2\pi/a)\times (\pm 0.5,\pm 0.5)$. This peak occurs in the other two materials as well (not shown) and its position is found to be independent on the doping level. Interestingly, ${\bf q}_{\pi,\pi}$ corresponds to the wavevector of the anti-ferromagnetic order observed in the parent compound. Because the Lindhard formula describes both spin and charge susceptibility, the predicted scattering enhancement around ${\bf q}_{\pi,\pi}$ is a precursor of the long-ranged spin order achieved in the absence of doping (Mott insulator). 

A second broad peak appears at ${\bf q}=(2\pi/a)\times (\pm 0.25,\pm 0.25)$. The exact position of this peak is material-dependent and ranges between $|q_x|=|q_y|=0.2$ and $|q_x|=|q_y|=0.3$ depending on the details of the band structure and the doping level, in analogy to the $q_y=0$ cut shown in Fig.~\ref{fig:lindhard}(b). Notably, this peak might easily escape experimental probes: due to its broadness, it seems to merge with the stronger peak at ${\bf q}_{\pi,\pi}$, especially if only the cut along the line $q_x=q_y$ is available. We will come back to this point in Sec.~\ref{sec:2d}. Finally, the third local maximum occurs at ${\bf q}=(\pm 0.25,\pm 0.1)$: the wavevector ${\bf q}=(2\pi/a)\times(\pm 0.25,0)$ is predicted to be a saddle point sitting between these local maxima. We note that a similar behavior was observed in a recent experiments by Thampy \etal \cite{thampy14}, who found sharp peaks at ${\bf q}=(2\pi/a)\times(0.25,\pm 0.015)$, separated by a saddle point at ${\bf q}=(2\pi/a)\times(0.25,0)$ \footnote{The quantitative discrepancy between our predictions and their finding may be attributed to either the different material ( ${\mathrm{La}}_{1.88}{\mathrm{Sr}}_{0.12}{\mathrm{CuO}}_{4}$), or the different technique (resonant elastic X-ray scattering). }.

\begin{figure}[b]
\centering	
\includegraphics[scale=0.5]{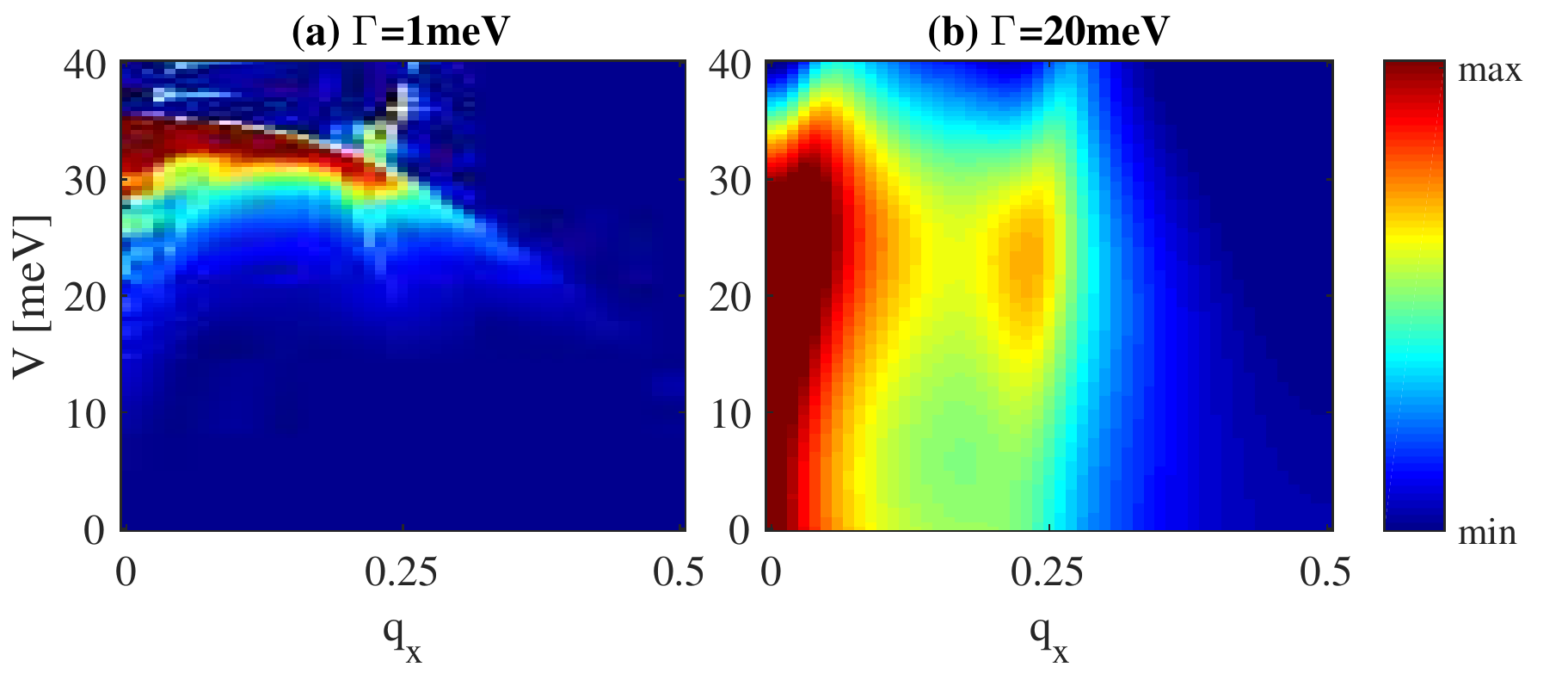}
\caption{STM spectra $g(q,\omega)$ along the line ${\bf q}=(q_x,0)\times2\pi$ for Bi2212 (see Table \ref{table1}) and $T=T^{dPW}$. A long quasiparticle's lifetime ((a) $\Gamma=1$meV) leads to dispersive peaks, while a short lifetime  ((b) $\Gamma=20$meV) leads to non-dispersive peaks. }\label{fig:figure2}
\end{figure}

\begin{figure*}[t]
\begin{tabular}{c c c c}	
\includegraphics[width=14cm]{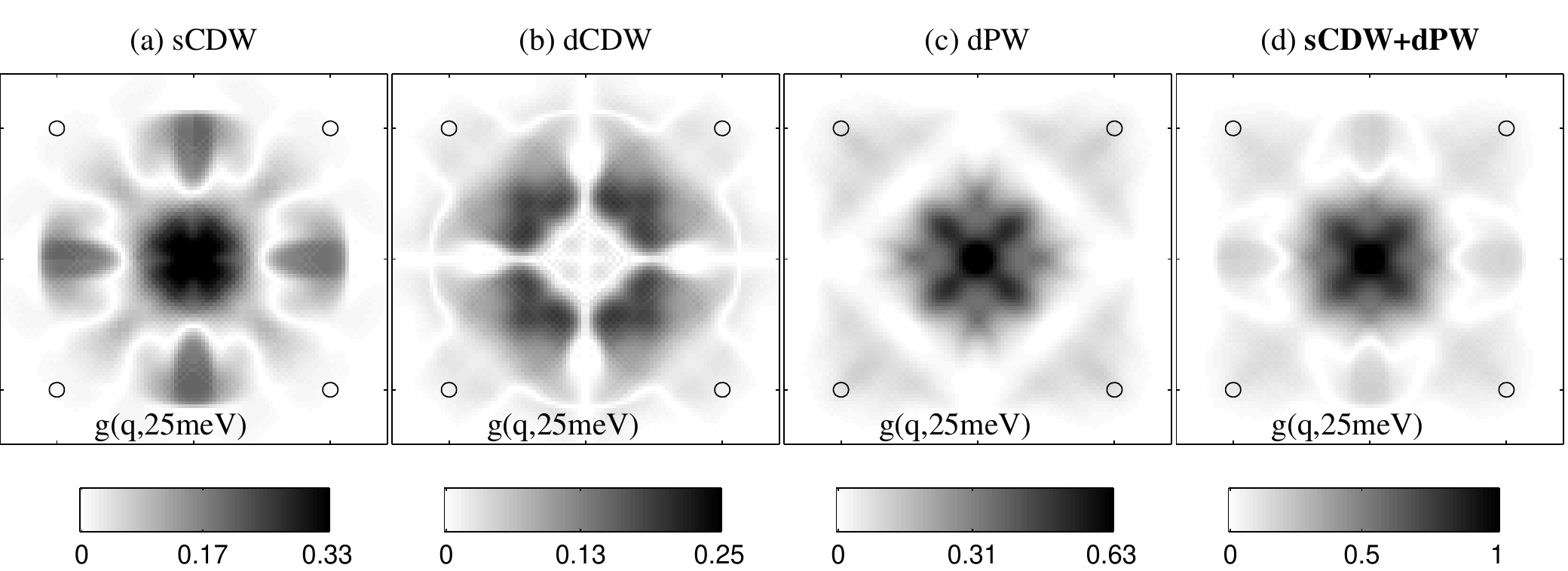} &
\includegraphics[scale=0.3]{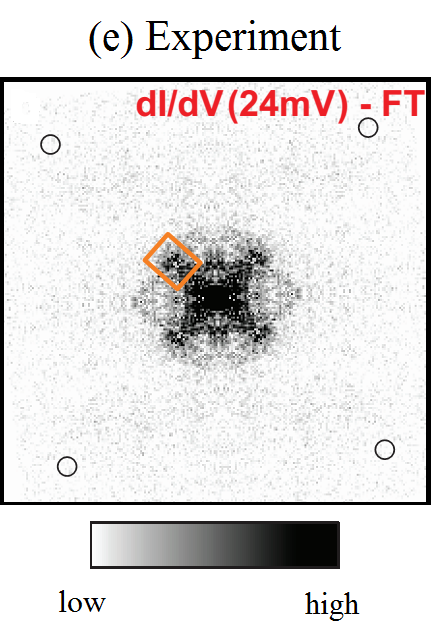}
\end{tabular}
\caption{(a-d) Numerical evaluation of Fourier-transformed STM spectra, Eq.~(\ref{eq:rhoqw2}), for different types of impurities:  local modulation of the chemical potential (sCDW), local modulations of the intra-unit-cell nematic order (dCDW), local modulations of the pairing gap (dPW). Model details: see column Bi2201 of Table \ref{table1}. The black circles denote the Bragg peaks at ${\bf q}=(\pm 2\pi,0)$ and $(0,\pm2\pi)$ . (e) Experimental measurement reproduced from Ref.~[\onlinecite{comin14}]. }\label{fig:allSTM}
\end{figure*}
\begin{figure*}[t]
\begin{tabular}{c c c c}	
\includegraphics[scale=0.85]{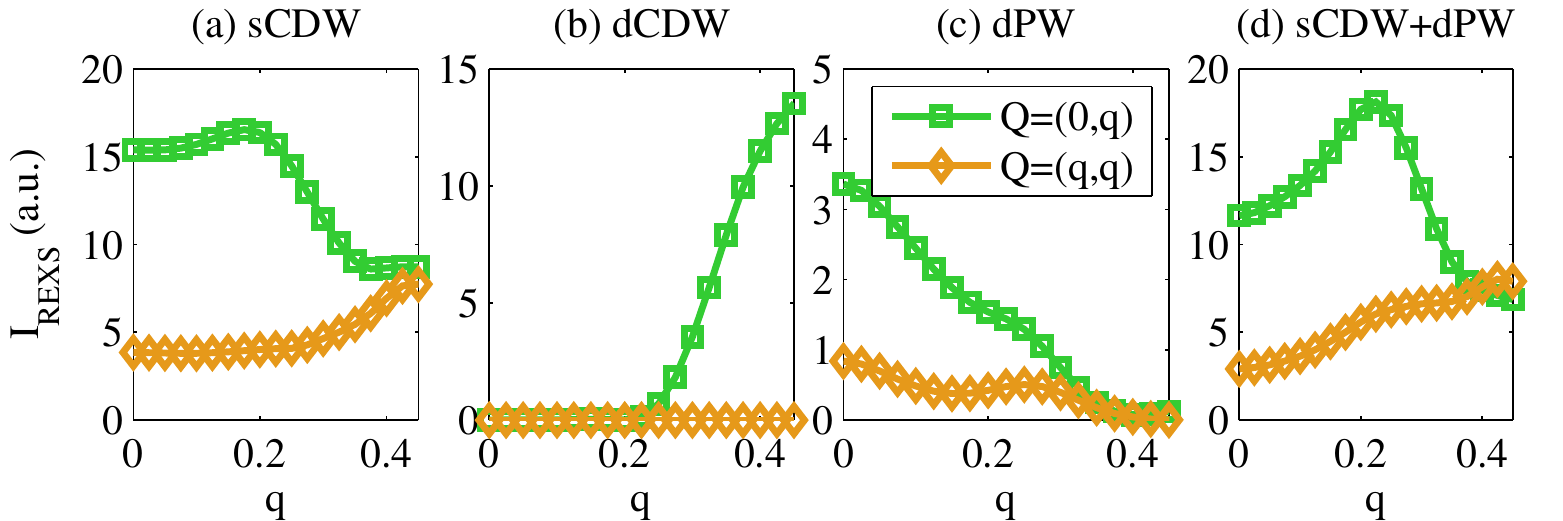} 
& \includegraphics[scale=0.3]{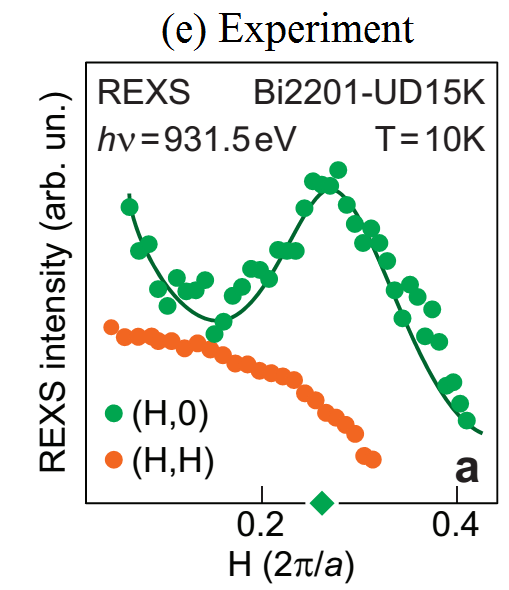}
\end{tabular}
\caption{(a-d) Numerical evaluation of the REXS signal, Eq.~(\ref{eq:REXS}), for different types of impurities (see caption of Fig.~\ref{fig:allSTM}). In order to easy the comparison with the experimental plot, the theoretical predictions in the $(q,q)$ direction (orange diamonds) have been reduced by a factor of 0.25 with respect to the $(q,0)$ direction (green squares). Model details: see column Bi2201 of Table \ref{table1} and $\Gamma_c=300$meV. (e) Experimental measurement reproduced from Ref.~[\onlinecite{comin14}]. The experimentally-observed peak at $q \approx 0.25$ in the $(q,0)$ direction (and its absence in the $(q,q)$ direction) is correctly reproduced by our simple model of Friedel oscillations.  }\label{fig:allREXS}
\end{figure*}

\subsection{STM: dispersive vs non-dispersive peaks}

We now proceed to discuss STM experiments, by first offering a brief summary of the main results of Ref.~[\onlinecite{dallatorre13}]. Specifically, in that paper we related the emergence of {\it non-dispersive} peaks in underdoped cuprates to their relatively large inverse quasiparticle lifetime $\Gamma$. In materials where $\Gamma$ is small (such as overdoped cuprates), the STM probe excites quasiparticles with an energy that precisely corresponds to the tip-sample voltage. In this case, energy and momentum conservation leads to the well-known ``octet model'' \cite{lee03}. This model predicts the emergence of seven inequivalent {\it dispersive} peaks, which can be found by connecting points on the Fermi surface where the pairing gap is equivalent to the tip-sample voltage. As shown for example by Nowadnick \etal \cite{nowadnick12}, these peaks are indeed reproduced by Eq.~(\ref{eq:rhoqw2}) in the limit of $\Gamma\to 0$. In contrast, for a finite $\Gamma$, the argument leading to the octet model does not apply because the quasiparticles' energy is not conserved. In this case, a numerical evaluation of Eq.~(\ref{eq:rhoqw2}) is necessary. As shown in Ref.~[\onlinecite{dallatorre13}] these calculations lead to non-dispersive peaks around the wavevectors connecting the antinodes. These scattering wavevectors are enhanced at all voltages for two reasons: (i) Any scattering is enhanced at the antinodes due to the Fermi surface nesting (in analogy to the analysis of Sec.~\ref{sec:lindhard}); (ii) The modulations of the pairing gap, $T^{dPW}_{\bf k}$ in Eq.~(\ref{eq:T3}), are proportional to the pairing gap $\Delta_{\bf k}$ and are therefore enhanced at the antinodes, where the latter is maximal.

The effect of $\Gamma$ on the calculated STM maps is highlighted in Fig.~\ref{fig:figure2}, where $\Gamma$ varies from 1meV (subplot (a)) to 20meV (subplot (b)), while all other parameters are kept fixed. The former plot displays dispersive peaks, while the latter mainly non-dispersive ones. Importantly, the temperature dependence of $\Gamma$ can explain the transition between dispersive peaks (at low temperatures) and non-dispersive peaks (at higher temperatures) reported in Ref.~[\onlinecite{yazdani10}].

\subsection{STM and X-ray: identifying the impurities}
\label{sec:2d}
To further clarify the nature of the main source of disorder (sCDW,dCDW, or dPW), we now consider the STM and REXS experiments of Comin \etal \cite{comin13}. Their two-dimensional Fourier-transformed STM signal is reproduced in Fig.~\ref{fig:allSTM}(e). The intensity of the signal is maximal in a  cross-shaped region, oriented in the $(\pm q,0)$ and $(0,\pm q)$ directions.  Figs.~\ref{fig:allSTM}(a-c)  represent our theoretical calculations for the three types of impurities defined in Eq.~(\ref{eq:T3}). The correct shape of the signal is reproduced only by local modulations of the pairing gap (dPW -- subplot (b)), suggesting that this is the dominant sources of disorder. A similar conclusion was reached by the independent analysis of Nunner \etal\cite{nunner2006fourier}. The experimental REXS measurement of the same material is reproduced in Fig.~\ref{fig:allREXS}(e). It shows a pronounced peak in the $(q,0)$ direction, and a monotonous behavior in the $(q,q)$ direction. A comparison with the theoretical curves, Fig.~\ref{fig:allREXS}(a-c), reveals that this effect is reproduced only by local modulations of the chemical potential (sCDW -- subplot (a)). 

This analysis leads to an apparent inconsistency: STM reveals local modulations of the d-wave pairing gap (dPW), while REXS reveals local modulations of the chemical potential (sCDW). The solution of this apparent paradox is hidden in the intrinsic properties of the two probes: STM measurements refer to low voltages and probe the scattering of quasiparticles with small energy $E \lesssim \Delta_0 \approx 20{\rm meV}$. In contrast, REXS probes the scattering of quasiparticles with energy $E\lesssim \Gamma_c \approx 300 {\rm meV}$. Due to the coherence factors appearing in Eq.~(\ref{eq:rhoqw2}), quasiparticles at different energies are mainly affected by different sources of disorder: low-energy quasiparticles are mainly affected by modulations of the pairing gap, while high-energy quasiparticles are mainly affected by modulations of the chemical potential (see also SI-3 of Ref.~[\onlinecite{dallatorre13}]). This effect becomes evident in the present calculation: the intensity of the STM signal is significantly stronger for dPW (Fig.~\ref{fig:allSTM}(c)) than for sCDW (Fig.~\ref{fig:allSTM}(a)), while the intensity of REXS is stronger for sCDW (Fig.~\ref{fig:REXS}(a)) than for dPW (Fig.~\ref{fig:REXS}(c)). The experimental results are then best reproduced by a superposition of both types of modulations (Figs.~\ref{fig:allSTM}(d) and  \ref{fig:allREXS}(d)). This result is in line with Jeljkovic \etal\cite{zeljkovic12}, who found a strong correlation between local perturbations of the pairing gap and of the chemical potential (identified there as atypical oxygen vacancies). Notably, local modulations of the intra unit-cell nematic order (dCDW) are inconsistent with the $q$ dependence of both STM and REXS signals. 

\begin{figure}[t]
\centering
\includegraphics[scale=0.5]{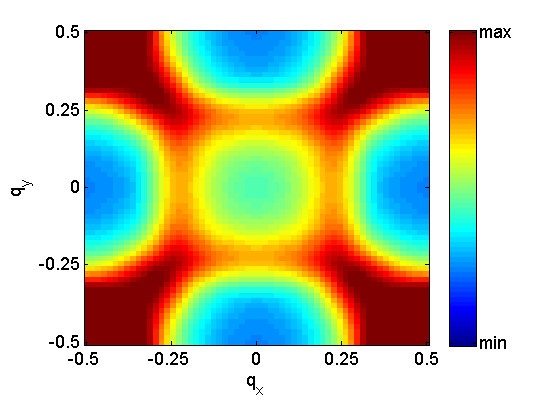} 
\caption{Two dimensional plot of the predicted REXS signal. Pronounced peaks are observed at ${\bf q}=(2\pi/a)\times(0.25,0)$ and ${\bf q}=(2\pi/a)\times(0.5,0.5)$. Model details: Bi2201 with sCDW+dPW impurities.}\label{fig:REXS2d}
\end{figure}

Let us now discuss a theoretical prediction made in Ref.~[\onlinecite{dallatorre13}], which appears to be in contradiction by the experiment of Comin\etal\cite{comin14}. Specifically, Ref.~[\onlinecite{dallatorre13}] predicted the existence of a peak in the REXS signal at wavevector $(0.25,0.25)$. In contrast, the experimental measurements of Ref.~[\onlinecite{comin14}] (orange curve in Fig.~\ref{fig:REXS}(e)) does not show any significant peak at $(0.25,0.25)$. We believe that the absence of the peak at $(0.25,0.25)$ is due to its blending with the larger and broader peak at $(0.5,0.5)$ along the same direction. \footnote{In the experimental curves reproduced in Fig.~\ref{fig:REXS}(e), the signal in $(0,q)$ and $(q,q)$ directions were normalized by a different multiplicative factor Indeed, if the same normalization had been used, the two curves should had converged in the $q\to0$ limit. As a consequence, the plot of Fig.~\ref{fig:REXS}(e) does not convey any information about the ratio between the $(0.25,0)$ and $(0.25,0.25)$ wavevectors, but only describes the behavior of the REXS intensity in the $(0,q)$ and $(q,q)$ direction, independently. } This phenomenon is clearly demonstrated in  Fig.~\ref{fig:REXS2d}, showing the predicted REXS intensity as a function of the two-dimensional wavevector ${\bf q}$. Note the close analogy with the results of the Lindhard susceptibility shown in Fig.~\ref{fig:lindhardB}.  We hope that future experiments will be able to confirm the hereby prediction of an increased scattering in the (q,q) direction. 

\begin{figure}
\includegraphics[scale=0.5]{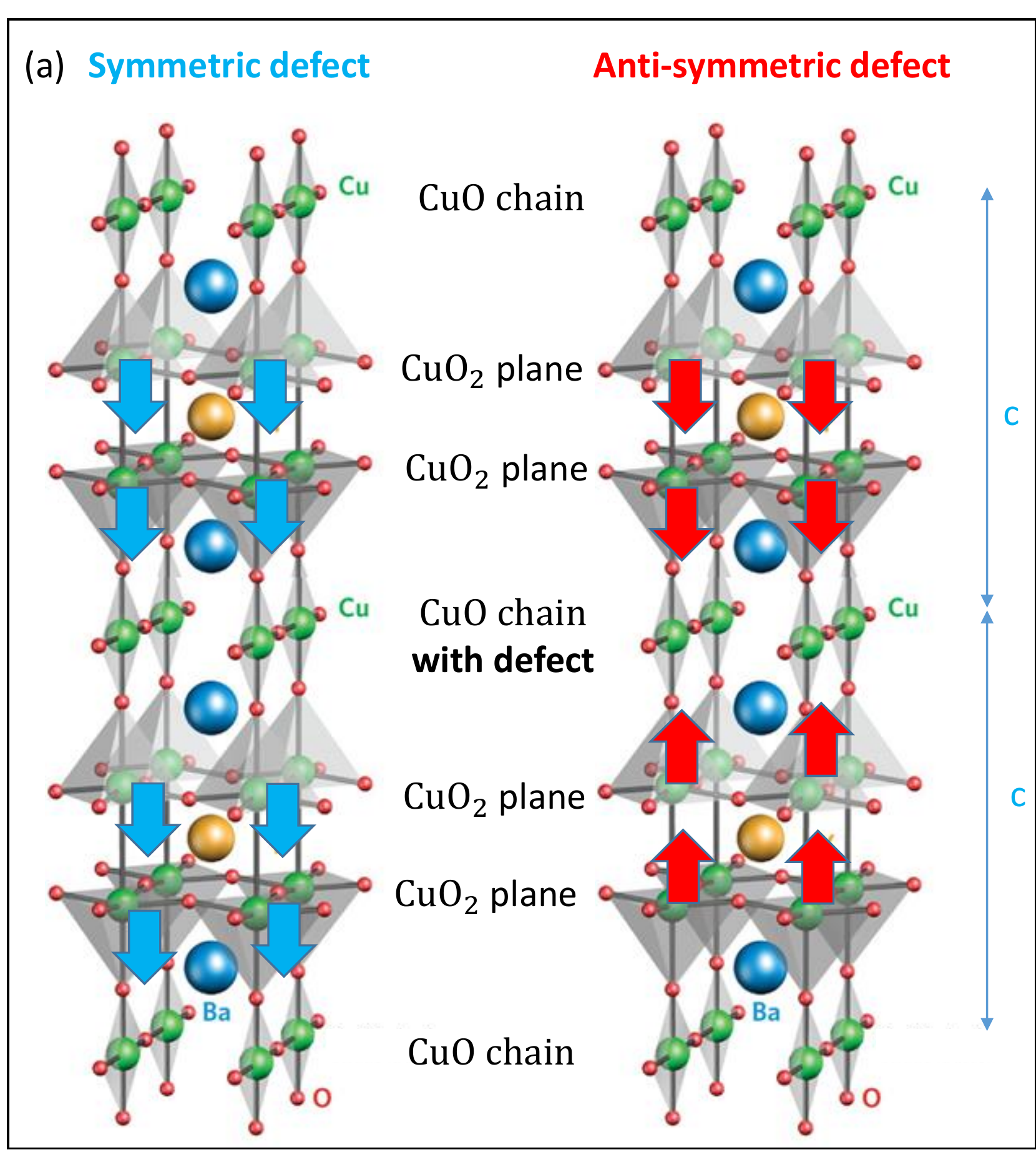}
\caption{Crystal structure of two neighboring unit cells of YBCO. A symmetric (anti-symmetric) defect located at a CuO chain induces the same (opposite) charge displacement on the CuO$_2$ planes of the two unit cells. Upward (downward) arrows indicate increased (decreased) charge density. graphical representation from Ref.~\onlinecite{barisic13}).}
\label{fig:caxisA}
\end{figure}

\begin{figure}
\centering
\includegraphics[scale=0.8]{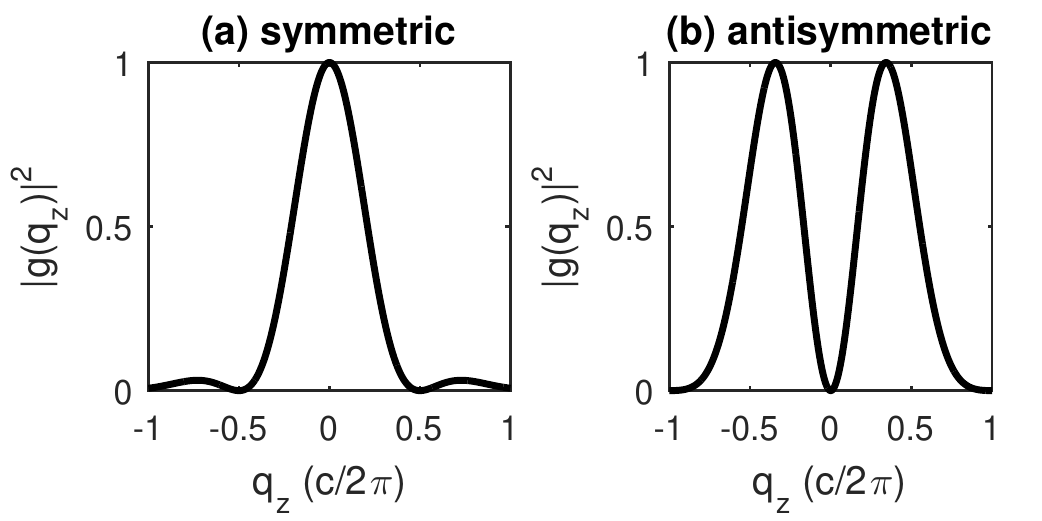}
\caption{Intensity of X-ray scattering as a function of the c-axis component of the wavevector $q_z$ for (a) symmetric and (b) anti-symmetric impurities, Eqs. (\ref{eq:gza}) and (\ref{eq:gza}). The latter curve is similar to the signal observed by Gerber \etal \cite{gerber15_three} at small magnetic fields and peaked around $q_z=\pm 1/2$.}
\label{fig:caxisBC}
\end{figure}

\subsection{X-ray: c-axis correlations}
Until this point we considered two-dimensional models only and analyzed correlations along the $a$ and $b$ principal directions only. Recently, Gerber \etal \cite{gerber15_three} found that the $c$-axis correlations provide a clear distinction between the short-ranged modulations observed at low magnetic fields and the long-range modulations found at large magnetic fields. Specifically, while the former is peaked around $k_z = 0.5 \times (2\pi/c)$ (or equivalently has a period of two unit cells), the latter is peaked at integer wave-vectors. As we will now explain, this observation is consistent with Friedel oscillations seeded by an impurity sitting at the interface between two unit cells. This situation is naturally realized in Y123, where the CuO chains are the main source of disorder and are equally spaces from the two neighboring CuO planes (see Ref.~[\onlinecite{alloul2009defects}] for a review).

If we neglect the tunnelling of electrons in the $c$ direction (i.e. among planes belonging to distinct unit cells), we obtain electronic bands that do not disperse in this direction. If we additionally assume that the scattering matrix $T$ and the Wannier function $W$ are separable functions of the spatial coordinates, the $q_z$ dependence can be factored out from Eqs.~(\ref{eq:grw}), (\ref{eq:rhoqw2}) and (\ref{eq:REXS}), leading to $I_{\rm REXS}({\bf q}) = |g(q_z)|^2 I_{\rm REXS}(k_x,k_y,\omega)$, where
\be g_z(q_z) = \sum_{\bf k_z} W^*_{k_z}T(k_z,q_z) W_{k_z+q_z}\label{eq:gz}\ee
and $I_{\rm REXS}(k_x,k_y,\omega)$ has been computed in the previous sections. Eq.~(\ref{eq:gz}) demonstrates that a non-trivial $q_z$ dependence can be obtained due to the shapes of the impurity and of the Wannier functions of the conduction-band electrons. The same approach goes through for hard X-ray diffraction, where one finds $\chi({\bf q}) = g(q_z) \chi(k_x,k_y,\omega)$, with $\chi(k_x,k_y,\omega)$ given by Eq.~(\ref{fig:lindhard}) (see Eq.~\ref{eq:Xray} in Appendix \ref{app:REXS2}).

For simplicity we now focus on an impurities that act (locally) on two neighboring CuO planes only:
\be T^{\pm}_{i} = \delta_{i,0} \pm \delta_{i,1} \ee 
Here $+$ ($-$) refers to a symmetric (antisymmetric) impurity, the index $i$ runs over the CuO layers, and $c$ is the lattice vector in the $z$ direction. A more accurate description should take into account the distinction between bonding and anti-bonding bands, but we defer this point to a future study. The effects of symmetric and anti-symmetric impurities are schematically plotted in Fig.~\ref{fig:caxisA}. 
The Fourier-transformed scattering amplitudes of symmetric and anti-symmetric impurities are then respectively given by
\begin{align}
T^{+}(k_z,q_z) &= 2\cos\left(\frac{c q_z}2\right)\;,\\
T^{-}(k_z,q_z) &= 2\sin\left(\frac{c q_z}2\right)\;.
\end{align}
Note that for symmetric impurities the intensity of $T$ is peaked at integer $k_z (c/2\pi) = 0,\pm 1,\pm 2,..$, while for antisymmetric impurities it is peaked at half integer $k_z(c/2\pi) = \pm 0.5, \pm 1.5, ...$. The intensities of the X-ray signals at low and high magnetic fields are therefore respectively consistent with anti-symmetric and symmetric impurities. To reproduce the experimental observations, we need to introduce trial Wannier functions. For simplicity we again refer to Gaussian function $W(z) = e^{- z^2/2c^2}$ which allow an analytic evaluations of Eq.~\ref{eq:gz}, leading to
\be
g_z(q_z)=\frac{\pi}{c^2} e^{- c^2 q_z^2/2} \cos^2\left(\frac{c q_z}2\right)
\label{eq:gzs}\ee
and
\be
g_z(q_z)=\frac{\pi}{c^2} e^{- c^2 q_z^2/2} \sin^2\left(\frac{c q_z}2\right)
\label{eq:gza}\;,\ee
respectively for symmetric and antisymmetric impurities. These curves are plotted in Fig.~\ref{fig:caxisBC}: subplot (b) reproduces the position and width of the experimentally observed signal\cite{gerber15_three}.

\section{Summary and outlook}
\label{sec:outlook}
In this paper we presented a theoretical modeling of  recent X-ray, REXS, and STM measurements of underdoped cuprates, with specific attention to Ghiringhelli \etal\cite{ghiringhelli12}, Chang \etal \cite{chang12}, Comin \etal \cite{comin14}. To interpret their experimental findings, these authors assumed the existence of a competing order, distinct from superconductivity, and associated with the spontaneous breaking of translational symmetry. The pseudogap energy scale could then correspond to the excitation gap required to restore the translational invariance. The association between the charge ordering and the pseudogap phase is however undermined by recent X-ray experiments revealing the same type of charge ordering in electron-doped cuprates\cite{daSilva15}, where a pseudogap phase is not expected to subsist. In addition, the experiment by Gerber \etal\cite{gerber15_three} showed that the oscillations observed at small magnetic fields have different $c$ axis correlations than those observed at large magnetic field\cite{gerber15_three}, indicating that these are two distinct effects. Similarly, recent measurements of the Hall conductivity \cite{grissonnanche2015onset} showing that the long-range ordered modulations appear only for magnetic fields that are larger than a critical value $H_c\approx 20 T$.


Following Ref.~[\onlinecite{dallatorre13}] we propose here that the modulations observed at small magnetic fields are simply due to Friedel oscillations around local sources of disorder. Because our interpretation is based on the Born approximation (first-order perturbation theory in the impurity strength) and we consider each scatterer independently, we expect the correlation length of the modulations to be independent on the concentration of impurities. This prediction has been now confirmed by two experimental observations: (i) Achkar \etal \cite{achkar13} modified the amount of disorder in Y123 through a thermal quench and observed that the correlation length of the observed oscillations was unchanged. (ii) The analysis of materials with similar band structure and different amount of intrinsic disorder (such as Bi2201\cite{comin13}, Bi2201\cite{yazdani14}, Bi2212\cite{shen14}, and Hg1201\cite{tabis14}) revealed an approximately constant correlation length. These findings are not consistent with theories of competing orders, in which the  predicted correlation length should be directly related to the amount of external disorder \cite{kivelson13}.  

By considering the scattering of short-lived quasipartice from local impurities, we can quantitatively reproduce all the experimental findings: Our model correctly predicts the wavevector and correlation length of the spatial modulations that were observed in X-ray (Figs. \ref{fig:lindhard}(b) and \ref{fig:lindhardB}), STM (Fig.~\ref{fig:figure2}), and REXS (Fig.~\ref{fig:allREXS}) experiments. The wavevector is similar (but not identical) to the distance between adjacent antinodes, where the Fermi surface is often quite nested. Our approach reproduces experimental observations that were interpreted as evidence for the d-wave symmetry of the oscillations (Figs.~\ref{fig:allREXS} - see also Ref.~[\onlinecite{dallatorre16}] for an in-depth analysis of the phase correlations observed by Fujita \etal \cite{fujita14}). Finally, it naturally accounts for the non-trivial $c$-axis correlations observed in X-ray experiments (Fig.~\ref{fig:caxisBC}).

To reproduce the experimental results, we introduced different models of local impurities and found that the most dominant type corresponds to local modulations of the chemical potential and of the pairing gap. In STM maps, the former contribution is generically dominant along the (q,q) direction, while the latter is dominant along the (0,q) direction (see Fig.\ref{fig:allSTM}). The interplay between these two sources of disorder connects to the earlier analysis of STM data in the presence of a magnetic field performed by Hanaguri \etal \cite{hanaguri09} and He \etal \cite{yanghe13}. These authors found that the ratio between the (q,q) and (0,q) components generically increases with magnetic field \footnote{Because the wavevector $(q,0)$ connects quasiparticles with opposite signs of the pairing gap, while the $(q,q)$ wavevector  connects quasiparticles with the same sign, these two contributions are often referred to as ``sign-reversing'' and ``sign-preserving''.}. 

\begin{figure}[h]
\includegraphics[scale=1]{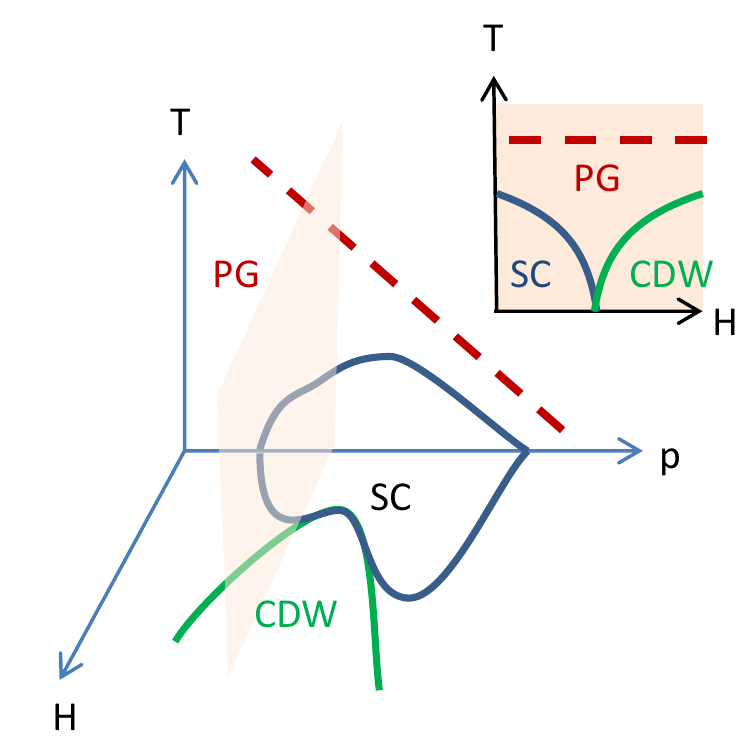}
\caption{Proposed phase diagram of BSCCO and YBCO compounds, following Ref.~[\onlinecite{wu14}]: the long-range-ordered CDW phase observed at high magnetic field is distinct from the pseudogap (PG) phase, characterized by an incoherent pairing gap of preformed pairs. The inset refers to doping levels $p\approx0.1$, where a direct transition between the superconducting (SC) and CDW phases is observed \cite{wu14}.}
\label{fig:phasediagram}
\end{figure}
\begin{figure*}[t]
\begin{tabular}{c c c c}
(a) small H & (c) small T & (e) large p &(g) dominant direction  \\
\includegraphics[scale=0.35]{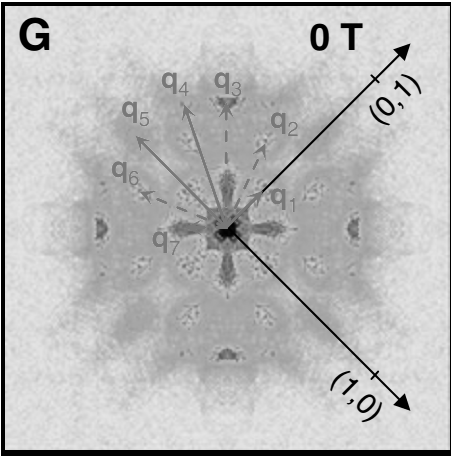}&
\includegraphics[scale=0.33]{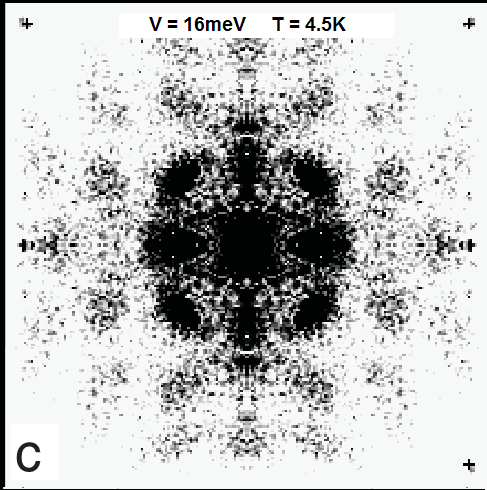}&
\includegraphics[scale=0.35]{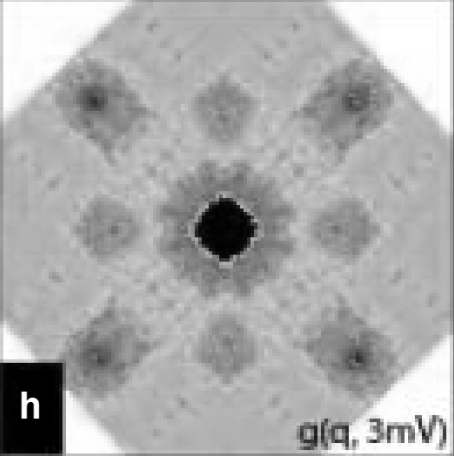}&
\includegraphics[scale=0.7]{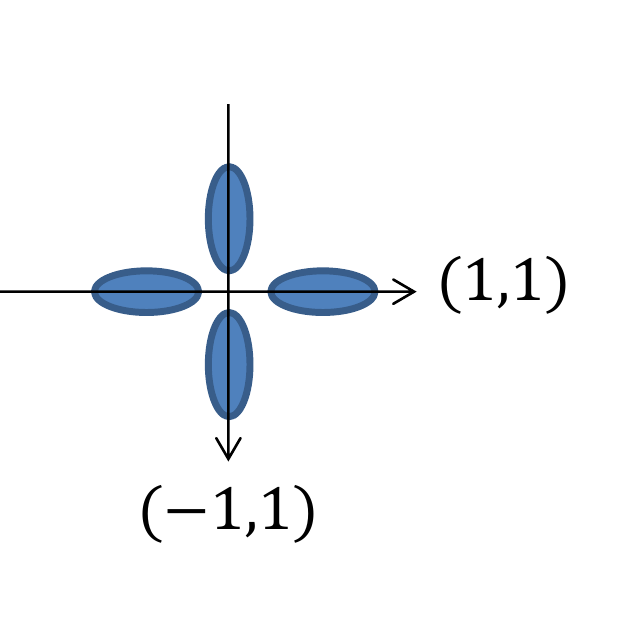}\\
(b) large H & (d) large T & (f) small p &(h) dominant direction \\
\includegraphics[scale=0.35]{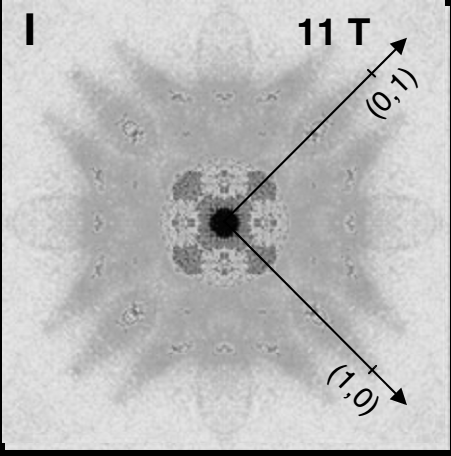}&
\includegraphics[scale=0.33]{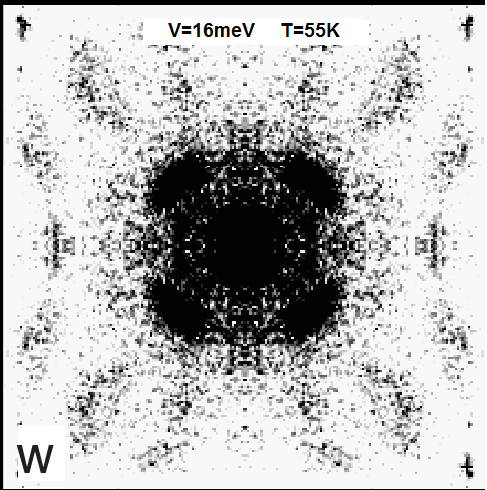}&
\includegraphics[scale=0.35]{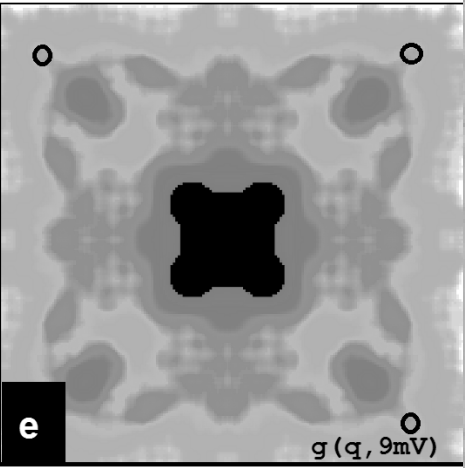}&
\includegraphics[scale=0.7]{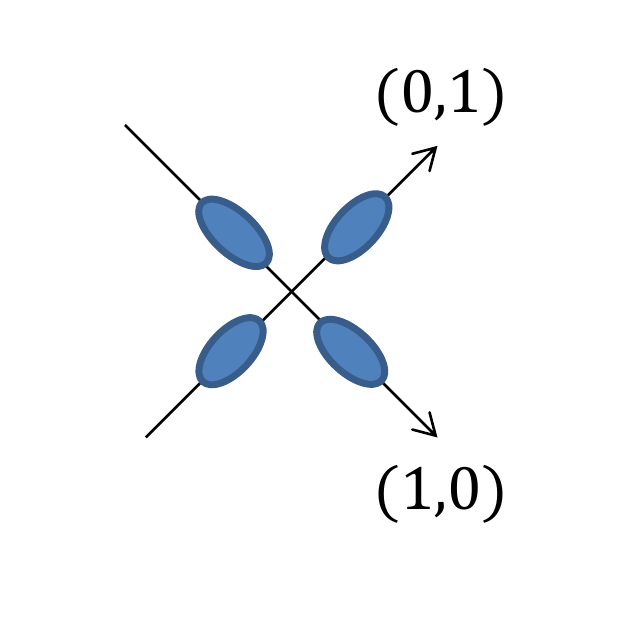}\\
\end{tabular}
\caption{Fourier-transformed STM measurements of different materials: (a-b) Ca$_{2–x}$Na$_x$CuO$_2$Cl$_2$ ($T_c$=28K)  at low and high magnetic field. Reproduced from Ref.\cite{hanaguri09};  (c-d)  Bi2212 ($T_c=37$K) at low and high temperature. Reproduced from Ref. [\onlinecite{fujita11}]; (e-f) Pb-Bi2201 at large and small hole doping. Reproduced from Ref. [\onlinecite{yanghe13}. (d) Deep in the superconducting phase the main source of scattering is along the (1,1) and (1,-1) directions. (h) When approaching the pseudogap phase the scattering is mainly along the (0,1) and (1,0) directions, signaling the presence of phase inhomogeneities.}\label{fig:dominant}
\end{figure*}

This effect can be understood by noting that in type-II superconductors external magnetic fields generate isolated vortexes, in whose core the pairing gap is locally suppressed\cite{pereg03_theory,wu11,wu13}. 
Vortexes are then similar to other types of local impurities, and generate Friedel oscillations around their center. This effect was for example directly observed by Simonucci \etal \cite{simonucci13_temperature}, who found Friedel oscillations around magnetic vortexes in the self-consistent solution of the BCS equations in the presence of a vortex. 

When the density of vortexes reaches a critical value, they can depin from local defects and give rise to a long-range-ordered phase. Following the proposal of Wu \etal \cite{wu11,wu13}, we believe this effect to be responsible for the formation of a long-range-ordered phase at large magnetic fields that was observed by  quantum oscillations \cite{doiron07}, NMR \cite{wu11,wu13}, and sound velocity \cite{leboeuf13} experiments. Indeed, the measured critical field $\sim 20T$ corresponds to an average distance between vortices of $d=\sqrt{\phi_0/B}\sim 100$A, which is comparable with the correlation length of Friedel oscillations. The CDW phase observed in cuprates would then be analogous to the field-induced spin density waves (FISDW) observed for example in Bechgaard salts (see Ref.[\onlinecite{chaikin1996}] for a review). Because the correlation length of Friedel osciilations depends on the amount of nesting at the antinodes, it is natural to expect the magnetic phase to be enhanced around $p=0.1$, where the antinodes are maximally nested. The resulting phase diagram is plotted in Fig.~\ref{fig:phasediagram}, and highlights our claim that the long-range ordered CDW phase is distinct from the pseudogap (PG) phase observed at zero magnetic field.

We now discuss how to utilize STM maps to further compare the effects of magnetic fields, temperature, and doping. Fig.~\ref{fig:dominant} shows that approaching the pseduogap phase (by increasing the temperature, or decreasing the doping) generically leads to an increase of scattering in the $(0,q)$ direction, which is associated with local modulations of the pairing gap. This observation is in agreement with recent muon spin rotation\cite{sonier13} ($\mu$SR) and NMR \cite{wu14} experiments, which detected enhanced static inhomogeneities in the pseudogap phase. A similar conclusion was reached in Ref.~[\onlinecite{cren00}], where the effects  of disorder were found to be similar to the effects of temperature and magnetic fields (see also Ref.~[\onlinecite{sacepe11}], where a pseudogap phase was found in disordered  thin films).  Inhomogeneities of the pairing gap are naturally accompanied by a reduction of the long-range coherence: the transition to the pseudogap phase may be due to a loss of coherence of the pairing gap \cite{emery95}, rather than to its disappearance. 

From the prospective of fermionic quasiparticles, the transition to the pseudogap phase is generically associated with an increase of the inverse lifetime $\Gamma$. The role of this quantity on ARPES measurement is well known and offers a simple explanation for the ``Fermi arcs'' observed in underdoped cuprates \cite{kanigel07,dessau12,dallatorre13}. STM \cite{alldredge08,dallatorre13} and transport \cite{hussey11} measurements show that the inverse quasiparticle lifetime $\Gamma$ is strongly enhanced in underdoped cuprates and probably diverges at the transition to the Mott insulator. This observation suggests a possible relation between $\Gamma$ and the critical temperature of cuprates \cite{dallatorre13}. Measuring the temperature dependence of $\Gamma$ would allow to distinguish between the effect of disorder (elastic scattering, which exists down to zero temperature) from the effects of interactions (inelastic scattering, which is supposed to diasppear at zero temperature). Surprisingly, although the inverse quasiparticle lifetime $\Gamma$ is commonly used in fitting virtually any experimental spectroscopic data, a systematic study of this quantity as a function of doping, temperature, and magnetic field has not been performed yet. We hope that the present work will motivate a new analysis of existing data along these lines.

\begin{acknowledgements}We acknowledge useful discussions with  Riccardo Comin, Andrea Damascelli, Giacomo Ghiringhelli, Nigel Hussey, Marc-Henri Julien, Steve Kivelson, and Max Metlistki, Giancarlo Strinati. ED acknowledges support from
Harvard-MIT CUA, NSF Grant No. DMR-1308435, AFOSR Quantum Simulation MURI, the ARO-MURI on Atomtronics, ARO MURI Quism program, Dr.~Max R\"ossler, the Walter Haefner Foundation, the Simons foundation, and the ETH Foundation. The work of EGDT and DD is supported by the Israel Science Foundation (grant No.1542/14). EGDT is thankful to the Aspen Center for Physics, where part of this work was performed during the winter-2014 workshop on ``Unconventional Order in Strongly Correlated Electron Systems''.  
\end{acknowledgements}

\appendix

\section{Technical details}
\subsection{Average over impurities}
\label{app:impurities}
In this section we consider the effects of several identical impurities, located at random positions. We find that the absolute value of $g({\bf q},\omega)$ is independent on  their position and therefore is an intrinsic property of the system. In the text, Eq.~(\ref{eq:gqw}), we assumed $g(r,\omega)$ to be symmetric under ${\bf r}\to-{\bf r}$. This approximation is valid in the presence of a single impurity located at the origin of the axis \footnote{Here we assume the impurity itself to be symmetric with respect to $r\to-r$. This assumption applies for example to impurities with $s$ and $d$ symmetry, but does not apply to impurities with $p$ symmetry}. To extend this treatment to systems with several impurities, we first notice that within the present framework (first order perturbation theory in the impurity strength), $g({\bf q\neq G},\omega)$ is given by a sum of terms, each referring to the scattering of quasiparticles from a single impurity: $g({\bf q},\w)=\sum_i g_i({\bf q},\w)$, where $i$ runs over all the impurities. To compute $g_i({\bf q},\w)$ we first consider a coordinate system whose origin is located at the center of the $i$th impurity, where Eq.~(\ref{eq:rhoqw2}) applies. We then shift back $g_i({\bf q},\w)$ to the common lab frame by the multiplication with $e^{i {\bf q}\cdot{\bf r}_i }$, and sum all the terms. In the case of $N$ identical impurities we obtain 
\be g({\bf q},\w) = \sum_{i=1}^N e^{i {\bf q}\cdot {\bf r}_i} g_0({\bf q},\w)\;.\label{eq:rhoqw3}\ee
Here $g_0({\bf q},\w)$ is the scattering amplitude from an isolated impurity located at the axis origin, computed from Eq.~(\ref{eq:rhoqw2}). In Eq.~(\ref{eq:rhoqw3}) the phase of $g({\bf q},\w)$ is determined by the random positions ${\bf r}_i$ and is therefore not predictable. In contrast, its absolute value averages to 
\be \langle |g({\bf q},\w)| \rangle = N g_0({\bf q},\w)\;. \ee
Here we used the observation that by definition, $g_0(q,\w)$ is a real function.
\subsection{Green's function approach to REXS}
\label{app:REXS}

In this section we derive Eq.~(\ref{eq:REXS}) within the Keldysh path-integral formalism. This approach allows us to extend the results of Abbamonte \etal \cite{abbamonte12} to finite temperatures. In REXS experiments X-rays are scattered upon the material to be examined at a frequency that allows the creation of a core hole, i.e. the excitation of an inner orbital of the atom to the conduction band (see Fig.~\ref{fig:REXS}). The action of the incoming field can be described as $V_{in}=E_{in}e^{i\w t}\delta(t) d\yd c + H.c. $, where $E_{in}>0$ describes the amplitude of the incoming x-rays, $\w$ its frequency, $d$ and $c$ are fermionic operators describing electrons (quasiparticles) respectively in the conduction band (in cuprates formed by $d$ orbitals) and in the core level. Shortly after, an electron from the conduction band fills in the core level and emits an X-ray photon, which is observed by the experimental setup. This decay process can be described by the operator $V_{out}(t) = a\yd_{out}e^{-i \w t} c\yd d$, where $a\yd_{out}$ creates an outgoing photon and $\lambda$ is the light-matter coupling. In perturbation theory, the outgoing field is given by
\be E_{out}(t) = \av{a_{out}	} \approx \av{c\yd d e^{-i \w t} + H.c.} \ee
where we assumed the initial state to be empty of outgoing photons. Applying perturbation theory (and neglecting oscillating terms), we obtain the Kubo formula
\bea 
E_{out}(t) = i E_{in}\Theta(t) \av{[d\yd(0) c(0),c\yd(t) d(t)]} \label{eq:nout}
\eea
where $\Theta(t)$ is the Heaviside theta function and $[...,...]$ is the commutation relation. In Keldysh notation, Eq. (\ref{eq:nout}) becomes the sum of eight terms with an odd number of ``classical'' fields $c,d$ and of ``quantum'' fields $\hat c,\hat d$. Four of these terms 
contain three quantum fields and their expectation values are identically equal to zero. We are then left only with terms containing three classical fields and one quantum field:
\begin{align} E_{out}(t)\nn =& \frac{i}2 E_{in}\Theta(t) \Big\langle\hat d^* (0) c(0) c^* (t) d(t) \\  &+ d^* (0) \hat c(0) c^* (t) d(t)+  d^* (0) c(0) \hat c^* (t) d(t) \nn \\& + d^* (0) c(0) c^* (t)  \hat d(t) \Big\rangle\label{eq:eight} \end{align}
For $t>0$ only the first two terms are non zero (and for $t<0$ the last two are non zero). Eq.~(\ref{eq:eight}) can be further simplified by introducing the retarded and Keldysh Green's functions $G^R_d=\av{d^*  \hat d}$ and $G^K_d=\av{ d^*  d}$: 
\be E_{out} =i  E_{in}\left[G^K_d(t)G^R_c(t) + G^K_c(t)G^R_d(t)\right]\label{eq:Eout2}\ee
Each of the two terms of Eq. (\ref{eq:Eout2}) corresponds to the product of two Greens functions, evaluated at the same time, or equivalently their convolution in the frequency domain:
\bea
E_{out}(\w) &=&\frac{i}2 E_{in} \int_{-\infty}^\infty d\w' \left[G^K_d(\w-\w')G^R_c(\w') \right.\nn\\&&\left.+ G^K_c(\w')G^R_d(\w-\w')\right]\label{eq:Eout2bis}.\label{eq:REXS3}\eea
At thermal equilibrium the Keldysh components satisfy the fluctuation-dissipation theorem $G^K_d(\w)=2{\rm Im}[G^R_d]\tanh(\w/2T)\approx 2{\rm Im}[G^R_d]{\rm sign}(\w)$ and $G^K_c=2{\rm Im}[G^R_c]\tanh((\w-E_h)/2T)\approx - 2{\rm Im}[G^R_c]$. In the limit of $T\to0$ we find
\bea
E_{out}(\w) &=& i E_{in} \int_{-\infty}^\infty d\w' \left[{\rm Im}[G^R_c](\w-\w')G^R_d(\w')\right. \nn\\&&+\left.{\rm sign}(\w') {\rm Im}[G^R_d](\w')G^R_c(\w-\w') \right]\label{eq:Eout3}.\label{eq:Eout4}\eea
The real component of $E_{out}$ (the component that is in phase with $E_{in}$)  has a particularly simple form
\be {\rm Re}[E_{out}](\w) = 2 E_{in}\int_{0}^\infty d\w'  {\rm Im}[G^R_c](\w-\w') {\rm Im}[G^R_d(\w')] \ee
Applying Karmers-Kronig relation we then obtain
\be E_{out}(\w) = 2 E_{in}\int_{0}^\infty d\w'  G^R_c(\w-\w') {\rm Im}[G^R_d(\w')] \ee
For a featureless core level with response function $G^R_c(q,\w)=[(\w+i\Gamma_c)]^{-1}$, we recover exactly the same expression as in Ref. [\onlinecite{abbamonte12}] and Eq.~(\ref{eq:REXS}) with $A=2E_{\rm in}$.

\subsection{From REXS to Lindhard}
\label{app:REXS2}

In the main text we provided an expression for the intensity of the REXS signal at zero temperature, Eq.~(\ref{eq:REXS}). Here we show that, in the case of non-resonant scattering from a Fermi gas this expression simply reduces to the Lindhard susceptibility (\ref{eq:lindhard}). The present derivation is a corollary of a more generic relation between the non-resonant limit of resonant inelastic scattering (RIXS) and density-density response functions (see Ref.~[\onlinecite{ament2011resonant}] for a review), and is brought here for completeness. Non-resonant scattering can be described as a REXS process in the limit of $\Gamma_c\to\infty$. In a Fermi gas with a local impurity, $G_0(k,\omega)=1/(\omega-\epsilon_k+i0^+)$, $W(k)=1$, and $T(k,k+q)=1$. Under these conditions Eq.~(\ref{eq:REXS}) and (\ref{eq:gqw}) give $I_{REXS} \to I_{\rm Xray} = |(A C({\bf q})/\Gamma_c|^2$ with
\begin{align} 
C({\bf q})&= \int_0^\infty d\omega' \sum_{\bf k}  {\rm Im}\left[\frac1{\omega' - \epsilon_{\bf k} + i0^+ }\frac1{\omega' - \epsilon_{\bf k+q} + i0^+}\right]\nonumber\\
&= \pi \int_0^\infty d\omega' \sum_{\bf k}  \frac{\delta(\omega'-\epsilon_{\bf k+q})}{\omega' - \epsilon_{\bf k}} + \frac{\delta(\omega'-\epsilon_{\bf k})}{\omega' - \epsilon_{\bf k+q} }\nonumber \\
&= \pi \int_0^\infty d\omega' \sum_{\bf k}  \frac{\delta(\omega'-\epsilon_{\bf k+q})}{\epsilon_{\bf k+q} - \epsilon_{\bf k}} + \frac{\delta(\omega'-\epsilon_{\bf k})}{\epsilon_{\bf k} - \epsilon_{\bf k+q} }\nonumber \\
& = \pi \sum_{\bf k}  \frac{n_{\bf k}}{\epsilon_{\bf k+q}- \epsilon_{\bf k}} + \frac{n_{\bf k+q}}{\epsilon_{\bf k} - \epsilon_{\bf k+q} }\;.
\end{align}
Here in the transition from the first to the second line we used $1/(x+i0^+)=1/x -i\pi \delta(x)$, and in the transition from the third to forth $\int_0^\infty \delta(\omega-\epsilon_k)=n_k$, where $n_k$ is the Fermi-Dirac distribution at $T=0$. We obtain 
\be I_{\rm Xray} = \left|\frac{A}{\Gamma_c}\sum_k \frac{n_{\bf k}-n_{\bf k+q}}{\epsilon_{\bf k}-\epsilon_{\bf k+q}}\right|^2 = \left|\frac{A}{\Gamma_c}\chi({\bf q})\right|^2\;.\ee
The present derivation can be extended to the case of non-trivial Wannier functions, and scattering amplitudes of the form $T({\bf k,k+q})=T_{\bf q}$ leading to
\be I_{\rm Xray} = \left|\frac{A}{\Gamma_c}\sum_{\bf k} W_{\bf k} T_{\bf q} W_{\bf k+q}^* \frac{n_{\bf k}-n_{\bf k+q}}{\epsilon_{\bf k}-\epsilon_{\bf k+q}}\right|^2 \label{eq:Xray}\ee

\subsection{Spin-orbit effects in REXS}
\label{app:spinorbit}
In this section we study the dependence of REXS scattering on the polarization of the incoming (i) and outgoing (o) photons. As an important result, we will show that in the absence of magnetic impurities, the intensity of the REXS signal is not affected by spin-orbit effects. For an isolated atom, the REXS intensity $I$ is given by the product of dipole matrix elements for the absorption and the emission:
\begin{equation}
I(\etahat_i,\etahat_o) \propto \left( \etahat^*_o \cdot \braketop{\psi_i}{\vr}{\psi_n} \right)
\left( \etahat_i \cdot \braketop{\psi_n}{\vr}{\psi_i} \right),
\end{equation}
where $\psi_i$ is the initial (and final) core electron state and $\psi_n$ is a valence $3d_{x^2-y^2}$ orbital with spin $\sigma$ at the same site.  In typical experiments one selects a resonance so that only $2p$ core levels with total angular momentum $j=3/2$ are excited to the valence band.  Thus the total polarization-dependent intensity is the sum over spin-orbit eigenstates $m_j=-3/2,-1/2,1/2,3/2$:
\begin{align}
I(\etahat_i,\etahat_o)  \propto & \sum_{m_j} \left( \etahat^*_o \cdot \braketop{2p^{3/2}_{m_j}}{\vr}{3d_{x^2-y^2},\sigma} \right)\nonumber\\
&\times
\left( \etahat_i \cdot \braketop{3d_{x^2-y^2},\sigma}{\vr}{2p^{3/2}_{m_j}} \right),
\end{align}
To compute the dipole matrix elements, we introduce a unit operator in the basis of separate spin and orbital angular-momentum eigenstates $\ket{m_\ell,m_s}$
\begin{align}
I\propto& \sum_{m_j,m_\ell,m_s,m^\prime_\ell,m^\prime_s} 
\left( \etahat^*_o \cdot \braket{2p^{3/2}_{m_j}}{m_\ell,m_s} \braketop{m_\ell,m_s} {\vr}{3d_{x^2-y^2},\sigma} \right)\nonumber\\
&\times \left( \etahat_i \cdot \braketop{3d_{x^2-y^2},\sigma}{\vr}{m^\prime_\ell,m^\prime_s} \braket{m^\prime_\ell,m^\prime_s}{2p^{3/2}_{m_j}} \right) \\
=& \sum_{m_j,m_\ell,m^\prime_\ell} 
\left( \etahat^*_o \cdot \braket{2p^{3/2}_{m_j}}{m_\ell,\sigma} \braketop{m_\ell} {\vr}{3d_{x^2-y^2}} \right)\nonumber\\
&\times \left( \etahat_i \cdot \braketop{3d_{x^2-y^2}}{\vr}{m^\prime_\ell} \braket{m^\prime_\ell,\sigma}{2p^{3/2}_{m_j}} \right) 
\end{align}
The Clebsch-Gordan matrix elements vanish unless $m_j= m_\ell+\sigma = m^\prime_\ell + \sigma$, and hence we require $m^\prime_\ell=m_\ell$.  We then have the further simplification:
\begin{align}
I\propto& 
\sum_{m_j,m_\ell}  \left| \braket{2p^{3/2}_{m_j}}{m_\ell,\sigma} \right|^2
\left( \etahat^*_o \cdot  \braketop{m_\ell} {\vr}{3d_{x^2-y^2}} \right)\nonumber\\
&\times\left( \etahat_i \cdot \braketop{3d_{x^2-y^2}}{\vr}{m_\ell} \right) 
\end{align}
If the Hamiltonian is spin-independent, i.e. if there is no spin-density wave, the amplitude is independent of the spin $\sigma$ of the photoelectron in the intermediate state and thus the two spins contribute equally to the coherent sum over histories, and we have
\begin{align}
I\propto& \sum_{m_\ell}   \left[ \sum_{m_j,\sigma}\left| \braket{2p^{3/2}_{m_j}}{m_\ell,\sigma} \right|^2 \right]
\left( \etahat^*_o \cdot  \braketop{m_\ell} {\vr}{3d_{x^2-y^2}} \right)\nonumber\\
&\times\left( \etahat_i \cdot \braketop{3d_{x^2-y^2}}{\vr}{m_\ell} \right) 
\end{align}
Now $\sum_{m_j,\sigma} \left| \braket{2p^{3/2}_{m_j}}{m_\ell,\sigma} \right|^2$ is the probability that a core electron with orbital angular momentum $m_\ell$ and unknown spin is in a total spin-$j=3/2$ state.  By spherical symmetry this is obviously independent of $m_\ell$, since $m_\ell$ is coordinate-dependent but $j$ is not.  Since this is an $m_\ell$-independent quantity, we obtain
\begin{align}
I\propto& 
\sum_{m_\ell}   
\left( \etahat^*_o \cdot  \braketop{m_\ell} {\vr}{3d_{x^2-y^2}} \right)
\left( \etahat_i \cdot \braketop{3d_{x^2-y^2}}{\vr}{m_\ell} \right)  \\
=& \braketop{3d_{x^2-y^2}}{ \left( \etahat_i \cdot \vr \right) \left( \etahat^*_o \cdot \vr \right)  }{3d_{x^2-y^2}}
\end{align}
We now have a tensorial matrix element that is not modulated by the spin-orbit effect except for the aforementioned constant prefactor that represents the contribution to resonant scattering only from $j=3/2$ core states.

\begin{figure}[b]
\centering
\includegraphics[scale=0.5]{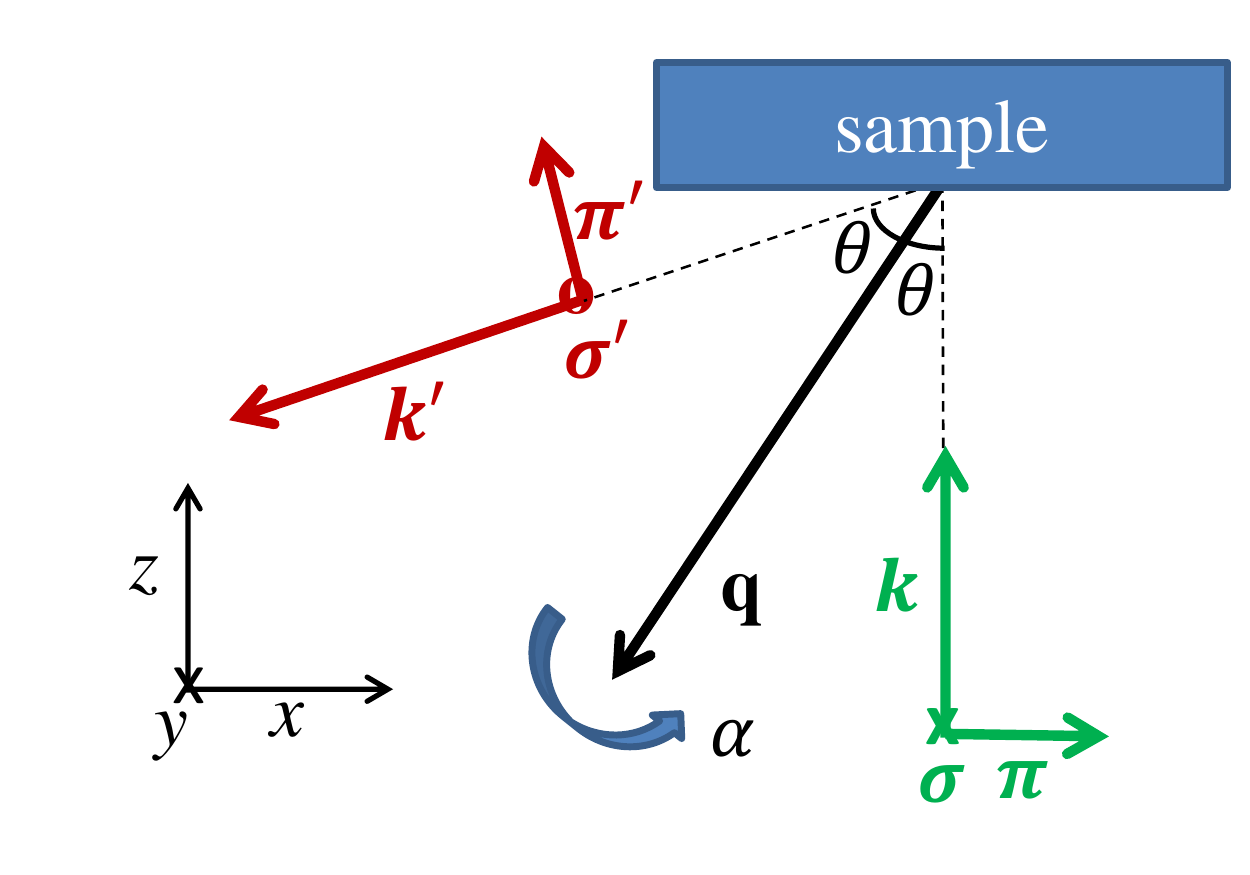} 
\caption{Experimental setup of Ref.~[\onlinecite{comin14}].}\label{fig:setup}
\end{figure}

\subsection{Polarization dependence of REXS}
\label{app:polarization}
\begin{figure}[t]
\centering
\includegraphics[scale=0.65]{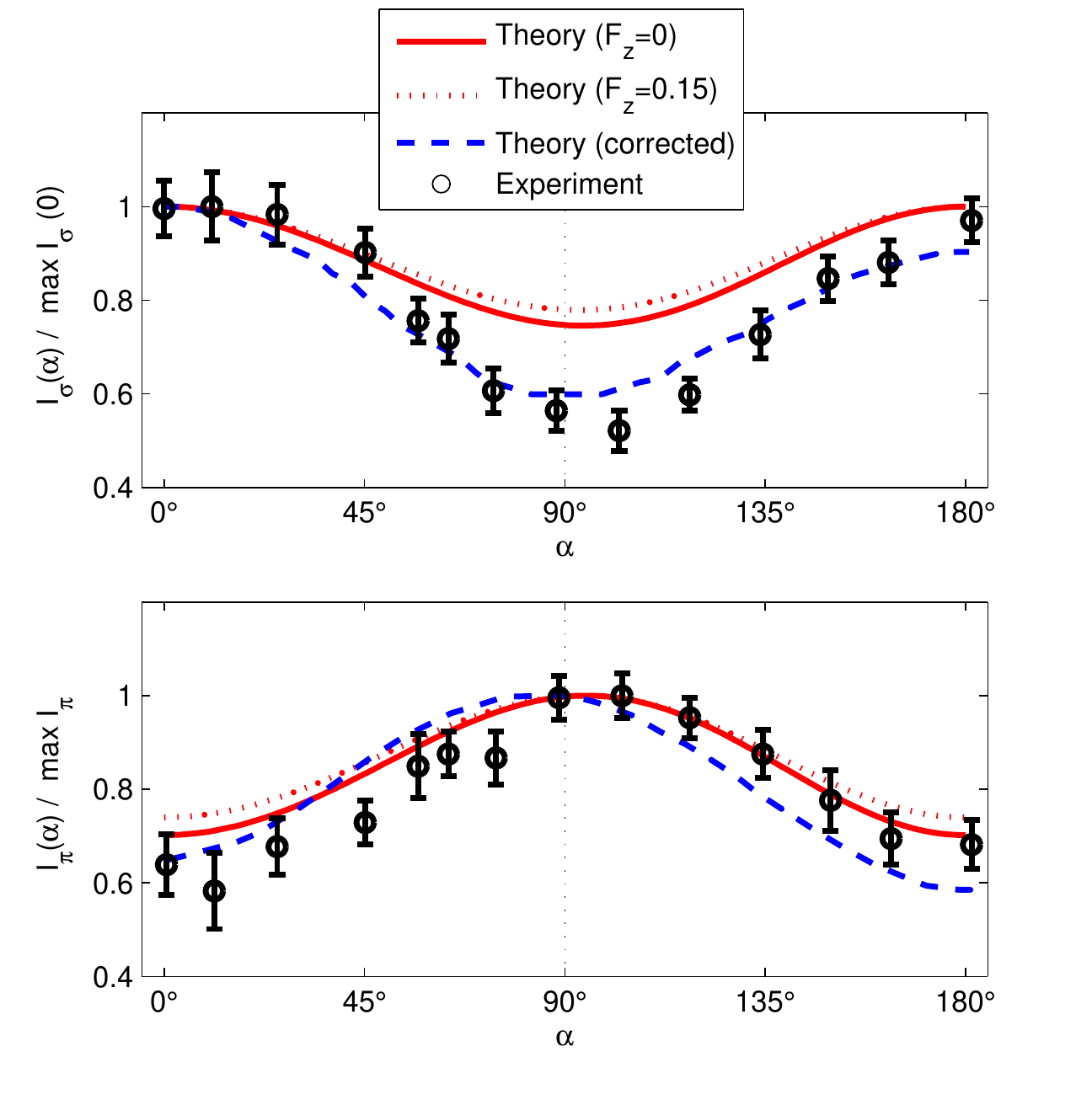} 
\caption{Polarization dependence -- comparison between theory and experiment. The continuous curves correspond to Eq.~(\ref{eq:polar1}) and Eq.~(\ref{eq:polar2}), obtained from Eq.~(\ref{eq:polar}) in the case of $F_z=0$. The dotted curves correspond to Eq.~(\ref{eq:polar}) in the case of $F_z=0.15$. The dashed curves are reproduced from Fig. S5 of Ref.[\onlinecite{comin14}] and include the corrections due to self-absorption. The errorbars indicate the experimental measurement of the Y675 sample reported in Ref.~[\onlinecite{comin14}].}\label{fig:polarization}
\end{figure}

In this section we study the dependence of the REXS signal on the wavevector of the incoming photon ${\bf k}$. This dependence was experimentally measured by Comin \etal \cite{comin14}, and used to identify the dominant type of charge modulations. According to the present single-band approach, the predicted ${\bf k}$ dependence is  instead identical for all types of modulations: Our model 
corresponds to the $s$-wave case considered by Comin \etal. This model is found to be in good agreement with the experimental measurements (see Fig.~\ref{fig:polarization}).

Our starting point is Eq.~(\ref{eq:polarization2}). Because the outgoing beam is not filtered according to its polarized, the measured signal is proportional to the sum of the intensities of the two outgoing polarizations: 
\be I_{\rm REXS} \propto \sum_{o=\sigma',\pi'} \Big | \etahat_o\cdot M \cdot \etahat_i ~\Big |^2 \label{eq:polar}\ee
where the tensor $M$ is defined by $M_{\alpha,\beta}=\bra{d}r_\alpha r_\beta\ket{d}$, $i=\sigma,\pi$ is the incoming polarization, and $o=\sigma',\pi'$ is the outgoing polarization. 
%
We denote by $F$ the diagonal matrix corresponding to $M$ in the principal axis of the lattice. 
Its three non-zero entries are $F_x=\bra{d}x^2\ket{d}$, $F_y=\bra{d}y^2\ket{d}$, and $F_z=\bra{d}z^2\ket{d}$. 
The ratio between these quantities was measured in Ref.~[\onlinecite{comin14}] and found to be
$F_z/F_x\approx F_z/F_y \approx 0.15$. The smallness of $F_z$ indicates that the conduction band has a small extension in the $z$ direction, in agreement with the theoretical calculations \cite{zhang88}, which predict a dominant $x^2-y^2$ character. In the experiment, the direction and the polarization of the incoming photons are kept fixed, while the wavevector ${\bf k}$ is modified by rotating the sample around the  vector ${\bf q}$ (Fig.~\ref{fig:setup}): the dipole matrix $M$ is then given by $M=R^T(\a) ~F~  R(\a)$, where the matrix $R$ represents a rotation of $\alpha$ degrees around the ${\bf q}$ axis. This expression corresponds to the predictions of Ref.~[\onlinecite{comin14}] for an $s$-wave modulation.

Without loss of generality we choose to work in Cartesian coordinates in which the incoming photon moves in the direction ${\bf k}=(0,0,1)$ with polarizations ${\bf \sigma}=(0,1,0)$ and ${\bf \pi}=(1,0,0)$.  As shown in  Fig.~\ref{fig:setup}, the polarization of the outgoing photon are ${\bf \sigma}'=(0,-1,0)$ and ${\bf \pi}'=(\cos2\theta,0,\sin2\theta)$, and ${\bf q} \equiv {\bf k}'-{\bf  k} \sim (\cos\theta,0,\sin\theta)$.
For $F_x=F_y=1, F_z=0$ we then find :
\begin{align}
I_\sigma(\a)&=\left(4\cos^3\frac{\a}2 \sin\frac{\a}2 \cos^2\theta\sin\theta\right)^2\nn\\
&+\left(\cos^2\a+\sin^2\a\sin^2\theta\right)^2\label{eq:polar1}
\\
I_\pi(\a)&=\left(4\cos\frac{\a}2 \sin^3\frac{\a}2 \cos^2\theta\sin\theta\right)^2\nn\\
&+\left(\cos^4\theta-\sin^2\theta(\sin^2\alpha+\cos^2\alpha\sin^2\theta)^2\right)^2
\label{eq:polar2}\end{align}
For $F_z\neq0$ one obtains more complex expressions, numerically depicted in Fig.~\ref{fig:polarization}. If we normalize $I_\sigma$ and $I_\pi$ by their maximal value (dotted curves of  Fig.~\ref{fig:polarization}), we find that ${\rm min}I_{\epsilon}/{\rm max} I_\epsilon \approx 0.75$ for both $\epsilon=\pi,\sigma$. The experimental measurements (symbols) instead show  ${\rm min}I_{\epsilon}/{\rm max}I_\epsilon\approx 0.6$. As explained by Comin \etal\cite{comin14}, this discrepancy can be attributed to the self-absorption effects \cite{achkar11}. The corrected signal (dashed curves) gives a good agreement between the predicted and measured ${\rm min}I_{\epsilon}/{\rm max} I_\epsilon$ ratios. 

Comin \etal additionally notice a significant asymmetry of the experimental data with respect to  $\alpha\to 180\degree-\alpha$. According to a specific statistical model, they attribute this discrepancy to a dominant $d$-wave character of the charge modulation. Their classification implies the realization of a multi-band model with inequivalent $O_x$ and $O_y$ orbitals, in contrast to the present analysis based on a single-band model. Interestingly, the statistical significance of their approach was not clearly established in the case of BSCCO. In addition, Fig.~\ref{fig:polarization} shows that the symmetry with respect to $\alpha\to 180\degree-\alpha$ is significantly modified by the corrections for self-absorption. We conjecture that the discrepancy between the present theoretical approach and the experimental measurements might be attributed to high-order corrections in the self-absorption, rather than to the symmetry of the charge modulations.


\end{document}